\documentclass[preprint, nobibnotes, aps, superscriptaddress]{revtex4-1}
\usepackage{amsmath}
\usepackage{amssymb}
\usepackage{hyperref}
\usepackage{physics}
\usepackage{graphicx}
\usepackage{color}

\renewcommand{\tensor}[1]{\overleftrightarrow{\boldsymbol{#1}}}

\begin{document}

\title{Designing Collective Non-local Responses of Metasurfaces}

\author{J. R. Capers}
\email{jrc232@exeter.ac.uk}
\affiliation{Department of Physics and Astronomy, University of Exeter, Stocker Road, Exeter, EX4 4QL}

\author{S. J. Boyes}
\affiliation{DSTL, Porton Down, Salisbury, Wiltshire, SP4 0JQ}

\author{A. P. Hibbins}
\affiliation{Department of Physics and Astronomy, University of Exeter, Stocker Road, Exeter, EX4 4QL}

\author{S. A. R. Horsley}
\affiliation{Department of Physics and Astronomy, University of Exeter, Stocker Road, Exeter, EX4 4QL}

\date{\today}
\begin{abstract}
    We propose a numerically efficient `adjoint' inverse design method to optimize a planar structure of dipole scatterers, to manipulate the radiation from an electric dipole emitter.  Several examples are presented: modification of the near-field to provide a 3 fold enhancement in power emission; re-structuring the far-field radiation pattern to exhibit chosen directivity; and the design of a discrete `Luneburg lens'.
    Additionally, we develop a clear physical interpretation of the optimized structure, by extracting `eigen-polarizabilities’ of the system.
    We find that large `eigen-polarizability’ corresponds to a large collective response of the scatterers.
    This framework may find utility in wavefront shaping as well as in the design and characterisation of non-local metasurfaces.
\end{abstract}

\maketitle

\section{Introduction}

Designing the scattering properties of materials is a fundamental challenge in a broad range of disciplines, from metamaterial design \cite{Quevedo-Teruel2019} to imaging through disordered media \cite{Mosk2012, Vellekoop2007, Vellekoop2008, Vellekoop2008a, Pendry2008}.
Metamaterials are structured at the sub-wavelength scale to control wave propagation, leading imaging beyond the limit set by diffraction \cite{Grbic2004}.
In recent years, there has been increasing interest in how appropriately designed metamaterials can induce virtually any desired wave effect, be that acoustic \cite{Ma2016} or electromagnetic \cite{Chen2016}.
To solve this problem the connection between incident and scattered fields must be established, before the metamaterial design can be engineered.
Great progress in mapping an input field to a desired output field has been made for imaging through disordered media.
However, this has involved manipulating the  incident wave itself,  rather than the scattering system \cite{Mosk2012, Vellekoop2007, Vellekoop2008, Vellekoop2008a, Pendry2008}.
In this work, we propose a semi-analytic framework for designing arbitrary scattering properties of systems of discrete scatterers.

Some of the earliest examples of metasurfaces are frequency selective surfaces \cite{patent, FrequencySelectiveSurfaces}.  
These are a class of periodically structured two-dimensional metal-dielectric structures designed to have specific reflection and transmission properties that depend upon the frequency of the incident wave.
These effects are facilitated by modifying the electromagnetic boundary condition through structuring of the surface to induce near-field and resonant effects.  These boundary conditions are typically both frequency and wave-vector dependent, and described by a complex effective surface impedance.
Frequency selective surfaces have since been used for many applications \cite{Anwar2018}, including as electromagnetic filters and as perfect absorbers \cite{Landy2008}.
Achieving perfect absorption in disordererd media has also attracted recent attention due to its potential use in several applications \cite{Pichler2019}, such as energy harvesting.
In this application, rather than designing surface impedance distributions, properties of the incident field are controlled, with gain and loss in the system being tuned.
However the aim of the two approaches is identical: arbitrary control of light.

By introducing spatial variation to the periodic basis of frequeny selective surfaces one may achieve inhomogeneous effective properties.  Inhomogeneity allows more abrupt phase shifts to be imparted upon the incident wave, providing additional degrees of freedom that can be exploited when manipulating the scattering of light \cite{Yu2011}.
This has led to the development of `metalenses', which can have superior bandwidth to traditional refractive lenses \cite{Chen2020, Lin2019}, as well as metasurface antennas \cite{Faenzi2019}.
Metasurface antennas have been designed to exhibit many valuable  properties, such as the opportunity to engineer bespoke beam-shaping, steering, polarization control and improved efficiency \cite{Faenzi2019}.
Instead of introducing inhomogeniety to a repeating basis, several classes of metasurface have been designed by `engineering disorder'.
This involves using an algorithm to selectively place scattering elements to form a metasurface with specific properties.
This principle has been used to design metasurface holograms \cite{Ni2013}, and for wavefront shaping \cite{Jang2018}.
What unites the seemingly disparate applications of holograms, metalenses, wavefront shaping and imaging through disorder is the problem of designing materials to realise a given wave effect.

The materials required for each of these functionalities can be designed using very similar methods.
For example, holograms \cite{Ni2013}, metalenses \cite{Chen2020} and beam shaping surfaces \cite{Faenzi2019} have all been designed using the Gercherg-Saxton algorithm \cite{Saxton1972}.
This method has been used extensively to find the maps of phase offset required to convert an incident plane wave into a given output.
Generalisations of this method to include control over the amplitude of the wave have also been developed \cite{Overvig2019}. 
However, this method neglects multiple scattering interactions and assumes that only a local phase offset is imparted upon the incident field.
This is inhibits the application of this design method to problems were non-local interactions are key, for example in achieving perfect anomalous reflection \cite{DazRubio2017, DazRubio2019}.
As well as this, the number of design degrees of freedom is reduced.

Due to broad demand for methods to design the scattering properties of materials, the problem of devising design methodologies has attracted recent attention \cite{Molesky2018}.
As well as the Gerchberg-Saxton, there are two other popular inverse design paradigms.
Firstly, geometry optimisation based upon the adjoint design method \cite{Giles2000} has been used to design many electromagnetic structures \cite{Lalau-Keraly2013,Molesky2018,Mignuzzi2019}.
Typically this procedure involves evaluating a cost function, which is to be extremised, over a given geometry using a full-wave solver.
Changes to the geometry are then made iteratively, so that the figure of merit is improved until a convergence is reached.
A key feature of the adjoint method is that the cost function contains both a `forward' and an `inverse' contribution \cite{Giles2000}.
Reciprocity \cite{LL2} is exploited to allow the `forward' and `inverse' parts to be calculated together, reducing the number of numerical simulations required to determine how material parameters should be changed.
In order to further simplify the numerical complexity axis symmetries \cite{Christiansen2020} and the locally periodic approximation \cite{Estourie2018}, which assumes a very sub-wavelength unit cell, are often exploited.
Secondly, machine learning \cite{Xu} and genetic algorithms \cite{Wiecha2018} have become extremely popular for solving the inverse design problem due to their ability to traverse large search spaces.  However while machine learning has a role to play in  optimisation processes, human intervention can provide more time-efficient design. It is the ambition of our current work to  seek a design method that  admits a clear physical interpretations of both the optimisation method and of the results, while being numerically efficient.  

In this work we present two contributions.
Leveraging the benefits of adjoint algorithms, we propose a semi-analytic framework to design the scattering properties of non-periodic arrangements of discrete dipolar scatterers.
Due to the efficiency of this method, we do not assume the weak scattering limit \cite{Gerke2010a}, which is often employed when it is assumed that multiple-scattering can be neglected \cite{Born1999}.
Instead, all interactions are taken into account so that all multiple-scattering effects are considered.
By examining these strongly non-local properties of the entire scattering system, we suggest an interpretation of the eigenvalues of the scattering system.
This provides explanatory detail on the mechanisms behind the optimisation procedure.

This paper is structured as follows.
In Section \ref{sec:solvingMaxwell} we briefly review the standard method used to solve Maxwell's equations in the presence of structures of dipolar scatterers.
Then, in Section \ref{sec:LDoS_responseMat} we discuss how one can use the Local Density of Optical States to characterise the effect of a photonic environment upon an emitter and propose an interesting interpretation of the eigenvalues of the response matrix.
In Section \ref{sec:DesigningScattering} we propose an iterative technique, based upon perturbation theory, to design the scattering properties of arrangements of scatterers.
To demonstrate how this approach may be applied, in Section \ref{sec:NumericalExamples} we illustrate the versatility of our technique with several numerical examples of manipulating the dipole field.  
We successfully demonstrate control over the near-field by enhancing power emission, over the far-field by re-shaping the angular distribution of the Poynting vector and we suggest how a discrete Luneburg lens might be realised by multiplexing such designs.

\section{Solving Maxwell's Equations for a System of Discrete Scatterers \label{sec:solvingMaxwell} }

We seek to design the scattering properties of an arrangement of magnetodielectric scatterers.  
Before proposing our solution to the inverse problem, we review the formulation of solutions to Maxwell's equations due to an arrangement of dielectric scatterers.

In general, the scatterering from a polarizible object is described by a summation over all the possible multipolar modes the object may possess \cite{Mie}.  We assume that only the electric and magnetic dipole modes are excited; this is consistent with several experimental observations \cite{Bohn2018, Vaskin2018}.
Formally, the dipole approximation is justified when the scatterers are sufficiently small (up to $\sim 1/k$, with $k$ being the wavenumber) and have separation $\geq 3r_0$ \cite{Savelev2014}, where $r_0$ is the radius of the scatterer.

We begin from Maxwell's equations for monochromatic waves of frequency $\omega$ in a general linear medium characterised by permittivity $\varepsilon$ and permeability $\mu$,
\begin{align}
    \boldsymbol{\nabla} \times \boldsymbol{\nabla} \times \boldsymbol{E} - k^2 \boldsymbol{E} = i \omega \mu \boldsymbol{J} + \omega^2 \mu \boldsymbol{P} + i \omega \mu \boldsymbol{\nabla} \times \boldsymbol{M}, \label{eq:helmholtzE}\\ 
    \boldsymbol{\nabla} \times \boldsymbol{\nabla} \times \boldsymbol{H} - k^2 \boldsymbol{H} = \boldsymbol{\nabla} \times \boldsymbol{J} + k^2 \boldsymbol{M} - i \omega \boldsymbol{\nabla} \times \boldsymbol{P} \label{eq:helmholtzH},
\end{align}
where $k = \omega \sqrt{\varepsilon \mu}$ is the wavenumber, $\boldsymbol{E}$ is the electric field, $\boldsymbol{H}$ is the magnetic field, $\boldsymbol{J}$ is the current density of the dipole emitter and $\boldsymbol{P}$ and $\boldsymbol{M}$ are polarization and magnetisation densities respectively.
Under the approximation that the scatterers and the source may be treated as points, the currents in Maxwell's equations (\ref{eq:helmholtzE}) and (\ref{eq:helmholtzH}) may be written explicitly as,
\begin{align}
    \boldsymbol{J} &= - i \omega \boldsymbol{\hat{p}} \delta (\boldsymbol{r} - \boldsymbol{r'}), &
    \begin{pmatrix}
        \boldsymbol{P} \\
        \boldsymbol{M}
    \end{pmatrix}
    &= 
    \tensor{\alpha}
        \begin{pmatrix}
        \boldsymbol{E} \\
        \boldsymbol{H}
    \end{pmatrix}
    \delta (\boldsymbol{r} - \boldsymbol{r}_n) ,
    \label{eq:dipoleSources}
\end{align}
where $\boldsymbol{\hat{p}}$ is the polarization of the source and
\begin{equation}
    \tensor{\alpha} = 
    \begin{pmatrix}
        \tensor{\alpha}_E & 0 \\
        0 & \tensor{\alpha}_H
    \end{pmatrix} ,
\end{equation}
is the polarizability tensor, in the absence of bianisotropy.
The location of the source is $\boldsymbol{r'}$ and of the $n^{\rm th}$ scatterer is $\boldsymbol{r}_n$.
It is important that the polarizability tensor obeys the usual requirements for energy conservation; namely that the energy the polarization/magnetization current absorbs from the incident field is equal to or greater than the energy re-radiated by the current.
This can be expressed as a constraint upon the elements of the polarizability tensor: ${\rm Im} [\tensor{\alpha}_{E,H}^{-1}] \geq -\mathbb{I} k^3 /(6 \pi \varepsilon)$ \cite{Belov2003, LL5, LL8}, where $\mathbb{I}$ is the unit tensor.

In general, any differential equation of the form 
\begin{equation}
    \boldsymbol{\nabla} \times \boldsymbol{\nabla} \times \tensor{G} (\boldsymbol{r}, \boldsymbol{r'}) - k^2 \tensor{G} (\boldsymbol{r}, \boldsymbol{r'}) = \delta (\boldsymbol{r} - \boldsymbol{r'}) \mathbb{I},
\end{equation}
may be solved in terms of the dyadic Green function \cite{Tai1993, NanoOps}
\begin{equation}
    \tensor{G} (\boldsymbol{r}, \boldsymbol{r'}) = \left( \mathbb{I} + \frac{1}{k^2} \boldsymbol{\nabla} \otimes \boldsymbol{\nabla} \right) \frac{e^{\pm ik|\boldsymbol{r} - \boldsymbol{r'}|} }{4 \pi |\boldsymbol{r} - \boldsymbol{r'}|},
    \label{eq:G}
\end{equation}
where $\mathbb{I} = {\rm diag}(1,1,1)$ is the unit tensor.
The two signs in the exponent correspond to advanced and retarded boundary conditions.
Integrating the Green function (\ref{eq:G}) against the source currents (\ref{eq:dipoleSources}), it can be shown that the solution to Maxwell's equations (\ref{eq:helmholtzE}) and (\ref{eq:helmholtzH}) can be written in terms of the outgoing wave solution and its curl $\tensor{G}_{EH} (\boldsymbol{r}, \boldsymbol{r'}) = \boldsymbol{\nabla} \times \tensor{G} (\boldsymbol{r}, \boldsymbol{r'})$ \cite{Sersic2011}.

To simplify notation and make it clear that our solutions are length-scale agnostic, we adopt a dimensionless unit system.  
We choose a characteristic length, $a$, by which to scale our coordinate system (for example, a natural choice might be $a = 1/k$).
In terms of this, we can define a unitless wavenumber $\xi = ka$.
Making use of the interchangeability of electric and magnetic fields, we work in units where free space impedance is one: $Z_0 = 1$.
In this unit system, after affecting the integration of the Green function (\ref{eq:G})  over the source currents (\ref{eq:dipoleSources}), the solution for the fields may be written as 
\begin{align}
    \begin{pmatrix}
        \boldsymbol{E}(\boldsymbol{r}) \\
        \boldsymbol{H}(\boldsymbol{r})
    \end{pmatrix}
    = 
    \begin{pmatrix}
        \boldsymbol{E}_s (\boldsymbol{r}) \\
        \boldsymbol{H}_s (\boldsymbol{r})
    \end{pmatrix}
    + \sum_n
    \begin{pmatrix}
        \xi^2 \tensor{G} (\boldsymbol{r}, \boldsymbol{r}_n) \tensor{\alpha}_E & i \xi \tensor{G}_{EH} (\boldsymbol{r}, \boldsymbol{r}_n) \tensor{\alpha}_H \\
        -i \xi \tensor{G}_{EH} (\boldsymbol{r}, \boldsymbol{r}_n) \tensor{\alpha}_E & \xi^2 \tensor{G} (\boldsymbol{r}, \boldsymbol{r}_n) \tensor{\alpha}_H
    \end{pmatrix}
    \begin{pmatrix}
        \boldsymbol{E} (\boldsymbol{r}_n) \\
        \boldsymbol{H} (\boldsymbol{r}_n) \\
    \end{pmatrix} ,
    \label{eq:fields}
\end{align}
where $\boldsymbol{E}_s (\boldsymbol{r})$ and $\boldsymbol{H}_s (\boldsymbol{r})$ are the source fields.
This is not yet a fully specified solution to Maxwell's equations, the total field applied at the location of each scatterer $(\boldsymbol{E}(\boldsymbol{r}_n), \boldsymbol{H}(\boldsymbol{r}_n))$ is not known.
These fields include two contributions: the source, and the scattering from all the other scatterers.
To determine the fields $(\boldsymbol{E}(\boldsymbol{r}_n), \boldsymbol{H}(\boldsymbol{r}_n))$, we must impose self-consistency upon the fields (\ref{eq:fields}).
To do this, we substitute $\boldsymbol{r} = \boldsymbol{r}_n$ into the field solutions (\ref{eq:fields}) and solve for $(\boldsymbol{E}(\boldsymbol{r}_n), \boldsymbol{H}(\boldsymbol{r}_n))$.
This yields the following matrix equation 
\begin{equation}
    \begin{pmatrix}
        \boldsymbol{E}(\boldsymbol{r}_1) \\
        \boldsymbol{H}(\boldsymbol{r}_1) \\
        \boldsymbol{E}(\boldsymbol{r}_2) \\
        \boldsymbol{H}(\boldsymbol{r}_2) \\
        \vdots 
    \end{pmatrix} 
            =
    \boldsymbol{R}^{-1}
    \begin{pmatrix}
        \boldsymbol{E}_s(\boldsymbol{r}_1) \\
        \boldsymbol{H}_s(\boldsymbol{r}_1) \\
        \boldsymbol{E}_s(\boldsymbol{r}_2) \\
        \boldsymbol{H}_s(\boldsymbol{r}_2) \\
        \vdots 
    \end{pmatrix},
    \label{eq:selfConsistency}
\end{equation}
where $^{-1}$ denotes matrix inversion, and $\boldsymbol{R}$ is the response matrix, defined as 

\begin{align}
    \boldsymbol{R} &= 
    \begin{pmatrix}
        \boldsymbol{R}_{11} & \boldsymbol{R}_{12} & \cdots \\
        \boldsymbol{R}_{21} & \boldsymbol{R}_{22} & \cdots \\
        \vdots & \vdots & \ddots 
    \end{pmatrix} , &
    \boldsymbol{R}_{ij} &= 
    \begin{pmatrix}
        \mathbb{I} \delta_{ij} - \xi^2 \tensor{\alpha}_E \tensor{G} (\boldsymbol{r}_i, \boldsymbol{r}_j) & -i \xi \tensor{\alpha}_H \tensor{G}_{EH} (\boldsymbol{r}_i, \boldsymbol{r}_j) \\
        i \xi \tensor{\alpha}_E \tensor{G}_{EH} (\boldsymbol{r}_i, \boldsymbol{r}_j) & \mathbb{I} \delta_{ij} - \xi^2 \tensor{\alpha}_H \tensor{G} (\boldsymbol{r}_i, \boldsymbol{r}_j)
    \end{pmatrix}
    \label{eq:Rdef}
\end{align}
The response matrix contains information about how each particle interacts with all of the other particles so that imposing this self-consistency condition is equivalent to solving the multiple-scattering problem for all $N$ scatterers at once.
It should be noted that this matrix inversion is potentially substantial, as $\boldsymbol{R} \in \mathbb{C}^{3 \times 2 \times N}$ (three spatial dimensions, for the magnetic and electric fields for each of the $N$ scatterers), making it the most numerically demanding part of our design process.
As this is a standard $\boldsymbol{A} \boldsymbol{x} = \boldsymbol{b}$ matrix problem, solving the self-consistency condition (\ref{eq:selfConsistency}) is easily facilitated numerically \cite{NumericalRecipes}.
Once the fields applied to each scatterer have been found in this way, they may be simply substituted into the expression for the full field (\ref{eq:fields}) so that the total fields for any structure may be calculated.

\section{The Local Density of Optical States and Eigenvalues of the Response Matrix \label{sec:LDoS_responseMat} } 

To quantify the effect of the scatterers upon the radiation of the source, we calculate the Partial (or Polarized) Local Density of Optical States (PLDoS).
For a fixed source current, this quantity is proportional to the power output \cite{Barnes2020, NanoOps}.
In terms of the total electric field, given by (\ref{eq:fields}), the PLDoS is given by \cite{Barnes2020}
\begin{equation}
    \rho (\boldsymbol{\hat{p}}, \boldsymbol{r}, \omega) = \frac{2 \epsilon_0 n^2}{\pi \omega} {\rm Im} [\boldsymbol{\hat{p}} \cdot \boldsymbol{E} (\boldsymbol{r}) ] .
    \label{eq:ldos}
\end{equation}
This quantity gives the number of electromagnetic modes per unit volume, for a given polarization $\boldsymbol{\hat{p}}$, position $\boldsymbol{r}$ and frequency $\omega$.
$\rho (\boldsymbol{\hat{p}}, \boldsymbol{r}, \omega)$ characterises both the changes to the radiation properties of the dipole emitter, and the modification of the electromagnetic modes at the emitter location.

To illustrate the physical meaning of the $\boldsymbol{R}$ matrix, we consider the simplest possible case: two scatterers with only an electric polarizability ($\tensor{\alpha}_H = 0$).
In this case, the response matrix connects the field induced by the source to the dipole moment induced in the scatterers as,  
\begin{equation}
    \begin{pmatrix}
        \boldsymbol{p} (\boldsymbol{r}_1) \\
        \boldsymbol{p} (\boldsymbol{r}_2)
    \end{pmatrix}
    = \tensor{\alpha}_E \cdot
    \begin{pmatrix}
        \mathbb{I} & -\xi^2 \tensor{G} (\boldsymbol{r}_1, \boldsymbol{r}_2) \cdot \tensor{\alpha}_E \\
        -\xi^2 \tensor{G} (\boldsymbol{r}_2, \boldsymbol{r}_1) \cdot \tensor{\alpha}_E & \mathbb{I}
    \end{pmatrix} ^{-1} \cdot 
    \begin{pmatrix}
        \boldsymbol{E}_s (\boldsymbol{r}_1) \\
        \boldsymbol{E}_s (\boldsymbol{r}_2)
    \end{pmatrix} .
    \label{eq:R_two_electric}
\end{equation}
The eigenvalues of $\boldsymbol{R}^{-1}$ satisfy the characteristic polynomial
\begin{equation}
    \boldsymbol{\lambda}^{-1} = \mathbb{I} \pm \frac{1}{2} \sqrt{4 \left[ \xi^2 \tensor{G} (\boldsymbol{r}_1, \boldsymbol{r_2}) \cdot \tensor{\alpha}_E \right] \cdot \left[ \xi^2 \tensor{G} (\boldsymbol{r}_2, \boldsymbol{r_1}) \cdot \tensor{\alpha}_E \right] },
    \label{eq:characteristic}
\end{equation}
which can be solved by making use of the reciprocity of the Green function \cite{Tai1993} to find that the eigenvalues of $\boldsymbol{R}$ are 
\begin{equation}
    \boldsymbol{\lambda} = {\rm diag} \left\{ \frac{\tensor{\alpha}_E}{\mathbb{I} \pm \xi^2 \tensor{G} (\boldsymbol{r}_1, \boldsymbol{r}_2) \cdot \tensor{\alpha}_E} \right\}.
    \label{eq:evals}
\end{equation}
One way to interpret the terms of (\ref{eq:characteristic}) are as multiple scattering events.
An interaction between the two electric dipole scatterers is comprised of scattering from $\boldsymbol{r}_1$ to $\boldsymbol{r}_2$ then back again.
This is what is expressed by the product of the Green functions in the discriminant of the characteristic equation (\ref{eq:characteristic}).
This understanding can be extended to more electric scatterers and scatterers with both electric and magnetic dipoles, whence more complex scattering processes become available.
However, this interpretation is still evident in the form of the polynomials.
Combining the fact that eigenvalues of $\boldsymbol{R}$ contain information about the collective response of the particles with (\ref{eq:evals}), the value of eigenvalue itself can be interpreted as the collective polarizability of the two scatterers.
Therefore, $\tensor{\alpha}_E^{-1} \cdot \boldsymbol{\lambda}$ gives the enhancement of the single particle polarizability due to the multiple scattering events between the two scatterers.
So, if $\tensor{\alpha}_E^{-1} \cdot \boldsymbol{\lambda} = 1$ there is no change to the single particle polarizability and multiple scattering events provide no enhancement.
On the other hand, a large $\tensor{\alpha}_E^{-1} \cdot \boldsymbol{\lambda}$ corresponds to a large enhancement to the response of a single scatterer, due to collective behaviour.

Additionally, the eigenmodes of $\boldsymbol{R}$ represent configurations of field that produce a certain collective response of the system.
As $\boldsymbol{R}$ has no symmetries these eigenmodes do not form an orthogonal basis \cite{Markel1995, Merchiers2007}, although the left and right eigenvectors of $\boldsymbol{R}$ do.
Using this left and right pair, the source field can be decomposed into the basis of these eigenvectors, $\boldsymbol{w}_n$ as 
\begin{equation}
    \boldsymbol{E}_s = \sum_{n=1}^{3N} c_n \boldsymbol{w}_n,
    \label{eq:expansion}
\end{equation}
where $N$ is the number of scatterers.
The expansion coefficient $c_n$ indicates which eigenmodes contribute most strongly to the response of the system.
Identifying these modes allows the response of the system to be understood and characterised by examining only a few eigenmodes, rather than the whole expansion (\ref{eq:expansion}).
The expansion coefficient is a useful tool in characterising the response of the system.
From this decomposition, one may find which modes are excited and how strongly so that the dominant response of the system may be isolated and examined.

Now that we have outlined how one can calculate, decompose and interpret the fields of a dipole emitter in the vicinity of dipole scatterers, under the assumptions of the point-dipole approximation, we shall proceed to apply a perturbative approach to \textit{design} the scattering properties of scatterer distributions.

\section{Designing Scattering Properties \label{sec:DesigningScattering} }

We calculate the figure of merit for the optimisation procedure as follows.
Firstly, the location of each scatterer and all of the fields are expanded to first order under a small perturbation to the location of each scatterer,
\begin{equation}
    \begin{aligned}
        \boldsymbol{r}_n &\rightarrow \boldsymbol{r}_n + \Delta \boldsymbol{r}_n, & \delta(\boldsymbol{r}-\boldsymbol{r}_n) &\rightarrow \delta (\boldsymbol{r}-\boldsymbol{r}_n) + \Delta \boldsymbol{r}_n \cdot \nabla \delta (\boldsymbol{r}-\boldsymbol{r}_n), \\
        \boldsymbol{E} &\rightarrow \boldsymbol{E}_0 + \boldsymbol{E}_1, & \boldsymbol{H} &\rightarrow \boldsymbol{H}_0 + \boldsymbol{H}_1.
    \end{aligned}
\end{equation}
Upon substituting these expressions into the solutions for the fields (\ref{eq:fields}), retaining only terms to first order, we obtain the following expressions for the corrections to the fields in terms of the unperturbed fields as
\begin{equation}
    \begin{aligned}
        \boldsymbol{E}_1 (\boldsymbol{r}) &= \zeta^2\left[ \tensor{G} (\boldsymbol{r}, \boldsymbol{r}_n) \tensor{\alpha}_E \boldsymbol{E}_{1} (\boldsymbol{r}_n) - \tensor{G} (\boldsymbol{r}, \boldsymbol{r}_n) \tensor{\alpha}_E \Delta \boldsymbol{r}_n \cdot \nabla \boldsymbol{E}_{0} (\boldsymbol{r}_n) \right] \\
        & + i \zeta \left[ \tensor{G}_{EH} (\boldsymbol{r}, \boldsymbol{r}_n) \tensor{\alpha}_H \boldsymbol{H}_{1} (\boldsymbol{r}_n) - \tensor{G}_{EH} (\boldsymbol{r}, \boldsymbol{r}_n) \tensor{\alpha}_H \Delta \boldsymbol{r}_n \cdot \nabla \boldsymbol{H}_{0} (\boldsymbol{r}_n) \right], \label{eq:correctionE}
    \end{aligned}
\end{equation}
\begin{equation}
    \begin{aligned}
        \boldsymbol{H}_1 (\boldsymbol{r}) &= \zeta^2 \left[ \tensor{G} (\boldsymbol{r}, \boldsymbol{r}_n) \tensor{\alpha}_H \boldsymbol{H}_{1} (\boldsymbol{r}_n) - \tensor{G} (\boldsymbol{r}, \boldsymbol{r}_n) \tensor{\alpha}_H \Delta \boldsymbol{r}_n \cdot \nabla \boldsymbol{H}_{0} (\boldsymbol{r}_n) \right] \\
        & - i \zeta \left[ \tensor{G}_{EH} (\boldsymbol{r}, \boldsymbol{r}_n) \tensor{\alpha}_E \boldsymbol{E}_{1} (\boldsymbol{r}_n) - \tensor{G}_{EH} (\boldsymbol{r}, \boldsymbol{r}_n) \tensor{\alpha}_E \Delta \boldsymbol{r}_n \cdot \nabla \boldsymbol{E}_{0} (\boldsymbol{r}_n) \right]. \label{eq:correctionH}
    \end{aligned}
\end{equation}
If one wishes to increase the power emission of the dipole, the figure of merit is $\propto {\rm Im} [\boldsymbol{E}_1 (\boldsymbol{r'})]$.
This is equivalent to increasing the PLDoS (\ref{eq:ldos}), producing more decay channels for the dipole emitter.
To structure the far-field, a point is the far field is chosen so that the figure of merit becomes $\propto {\rm Im} [\boldsymbol{E}_1 (\boldsymbol{r}_{\rm far field})]$, so that directivity in the direction of $\boldsymbol{r}_{\rm far field}$ is enhanced.

For each iteration $i$, the position of the $n^{\rm th}$ scatter is updated according to
\begin{equation}
    \Delta \boldsymbol{r}_n^{i+1} = \boldsymbol{r}_n^i + \Delta \boldsymbol{r}_n \times {\rm sign} \left[ {\rm Im} \left\{ \xi^2 \tensor{G} (\boldsymbol{r}, \boldsymbol{r}_n) \cdot \tensor{\alpha}_E \cdot \nabla \boldsymbol{E}_0 (\boldsymbol{r}_n) + i \xi \tensor{G}_{EH} (\boldsymbol{r}, \boldsymbol{r}_n) \cdot \tensor{\alpha}_H \cdot \nabla \boldsymbol{H}_0 (\boldsymbol{r}_n) \right\} \right].
    \label{eq:positionUpdate}
\end{equation}
In this way, the figure of merit may be iteratively increased.
It is clear that this is an adjoint calculation: the gradient operators act on the total field applied \cite{Bennett2019} to the scatterer located at $\boldsymbol{r}_n$, meaning that the effect of the change of the position of $\boldsymbol{r}_n$ upon the field at all of the other scatterers is implicitly included in the calculation.

A schematic of how our proposed optimisation procedure works is shown in Figure \ref{fig:setup}.
It is important to note the checks performed.
As the scatterers are modelled as points, it is necessary to ensure they remain properly separated so that the dipole approximation remains valid.
To guarantee this, scatterers are not allowed to move within a smoothing distance $d_0$ of each other.
It is also possible that there is no move the scatterer can take which improves the figure of merit, so these moves are also blocked.

To elucidate this process, we now present several examples of the application of the procedure outline in Figure \ref{fig:setup}.

\section{Numerical Examples \label{sec:NumericalExamples} }

\begin{figure}[h]
    \centering
    \includegraphics[width=\linewidth]{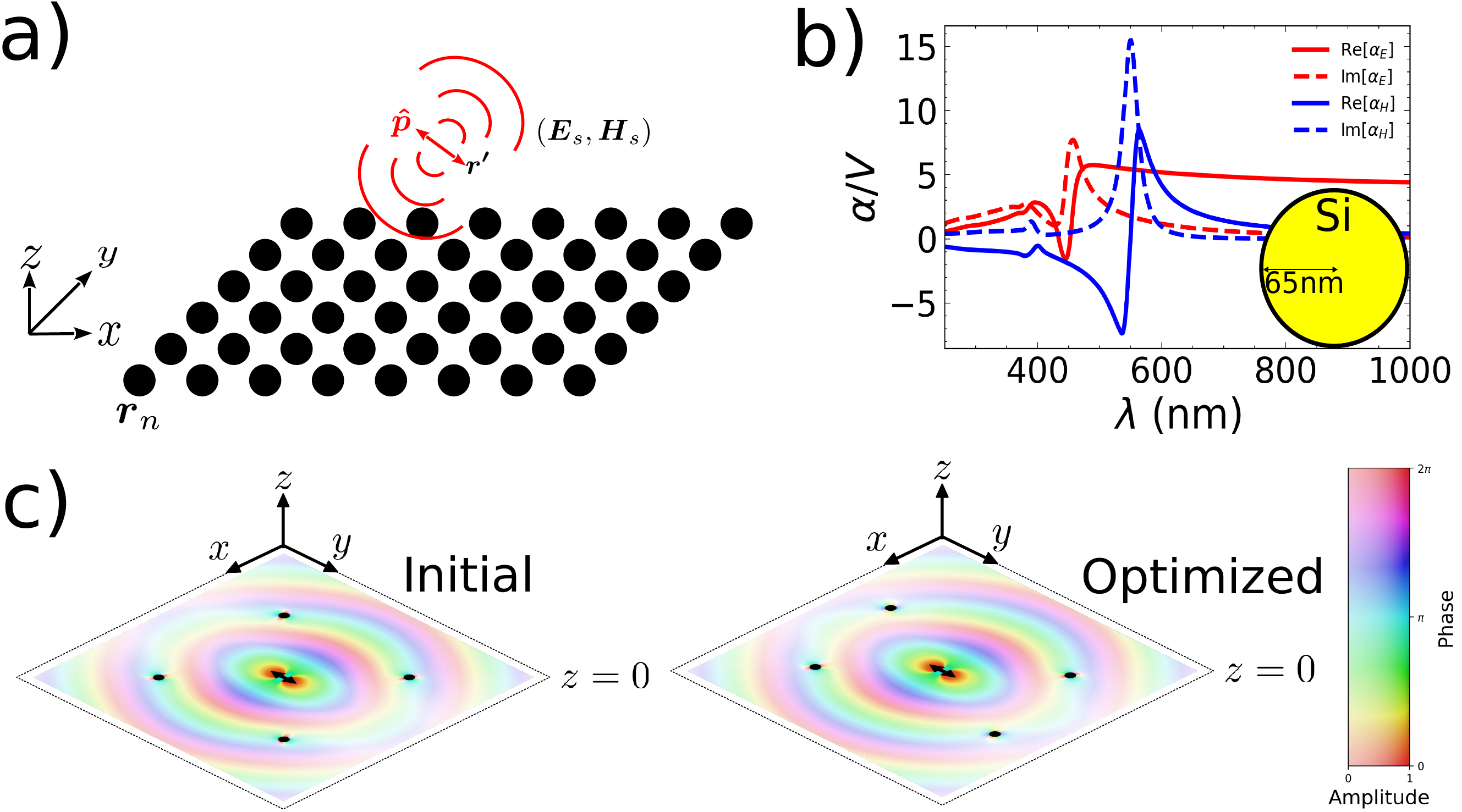}
    \caption{The situation considered in our numerical examples, showing (a) the numerical setup.  A dipole emitter with polarization $\boldsymbol{\hat{p}}$ is located at $\boldsymbol{r'}$.  This generates source fields ($\boldsymbol{E}_s$, $\boldsymbol{H}_s$), which are scattered by $N$ silicon spheres located at $\boldsymbol{r}_n$.  The scatters have optical properties shown in (b).  The scatterers are assumed to be isotropic silicon spheres of radius 65 nm, with the frequency dispersion of the polarizability extracted from experimental data \cite{Green2008}.
    A simple example of the design process is shown in (c).  In the initial configuration, scatterers are arranged in a square array.  This symmetry is broken by the optimization procedure to enhance power emission.}
    \label{fig:setup}
\end{figure}

To demonstrate the versatility and strengths of our proposed method, we apply it to solve several inverse design problems.
Here, we shall demonstrate the ability of our procedure to design the near field and far field equally, as well as how the results generated can be used to construct devices with more complex purposes.
The situation we consider is shown in Figure \ref{fig:setup}.
A dipole emitter at 550 nm with polarization $\hat{\boldsymbol{p}}$ is located at $\boldsymbol{r'}$.
Near this, we place a structure made from discrete scatterers.
These are modelled as isotropic silicon spheres of radius 65 nm, and polarizabilities calculated to match this.
A figure of merit is chosen and according to (\ref{eq:positionUpdate}) the locations of the scatterers are iteratively updated to improve this figure of merit.
This result of this procedure is shown schematically in Figure \ref{fig:setup}(c).
The optical properties of silicon have been extracted from experimental data \cite{Green2008}, then combined with the Mie $a_1$ and $b_1$ coefficients \cite{Bohren2004}, to allow the calculation of electric and magnetic polarizabilities \cite{Evlyukhin2010}.
Parameters used for all simulations are shown in Table \ref{tab:parameters}.
\begin{table}[h]
    \centering
    \begin{tabular}{c c c}
    \hline 
    \hline
    \textbf{Parameter} & \textbf{Description} & \textbf{Value} \\
    \hline 
       $r_0$ & Scatterer Radius & 65 nm \\
       $\lambda$ & Wavelength & 550 nm \\
       $\alpha_E / (\epsilon_0 V)$ & Electric Polarizability & 5.42 + $i$ 1.76  \\
       $\alpha_H / V$ & Magnetic Polarizability & 0.290 + $i$ 15.5 \\
       $d_0$ & Smoothing Length & 3$r$ = 195 nm \\
       $\Delta \boldsymbol{r}_n$ & Max. Step Size & 0.01 \\
       $\boldsymbol{r'}$ & Emitter location & (0, 0, 0) \\
    \hline
    \hline
    \end{tabular}
    \caption{Parameters used for all numerical calculations.}
    \label{tab:parameters}
\end{table}
It has been assumed that the scatterers are isotropic, so the electric and magnetic polarizability tensors have the form
\begin{align}
    \tensor{\alpha}_E &=
    \begin{pmatrix}
        1 & 0 & 0 \\
        0 & 1 & 0 \\
        0 & 0 & 1
    \end{pmatrix} \alpha_E, &
    \tensor{\alpha}_H &=
    \begin{pmatrix}
        1 & 0 & 0 \\
        0 & 1 & 0 \\
        0 & 0 & 1
    \end{pmatrix} \alpha_H .
\end{align}
The complex numbers $\alpha_E$ and $\alpha_H$ are calculated according to \cite{Bohren2004} as
\begin{align}
    \alpha_E &= i \frac{6 \pi}{k^3} a_1, & \alpha_H &= i \frac{6 \pi}{k^3} b_1,
\end{align}
where $a_1$ and $b_1$ are the Mie coefficients for the dipole modes of a spherical scatterer.

\begin{figure}[h]
    \centering
    \includegraphics[width=\linewidth]{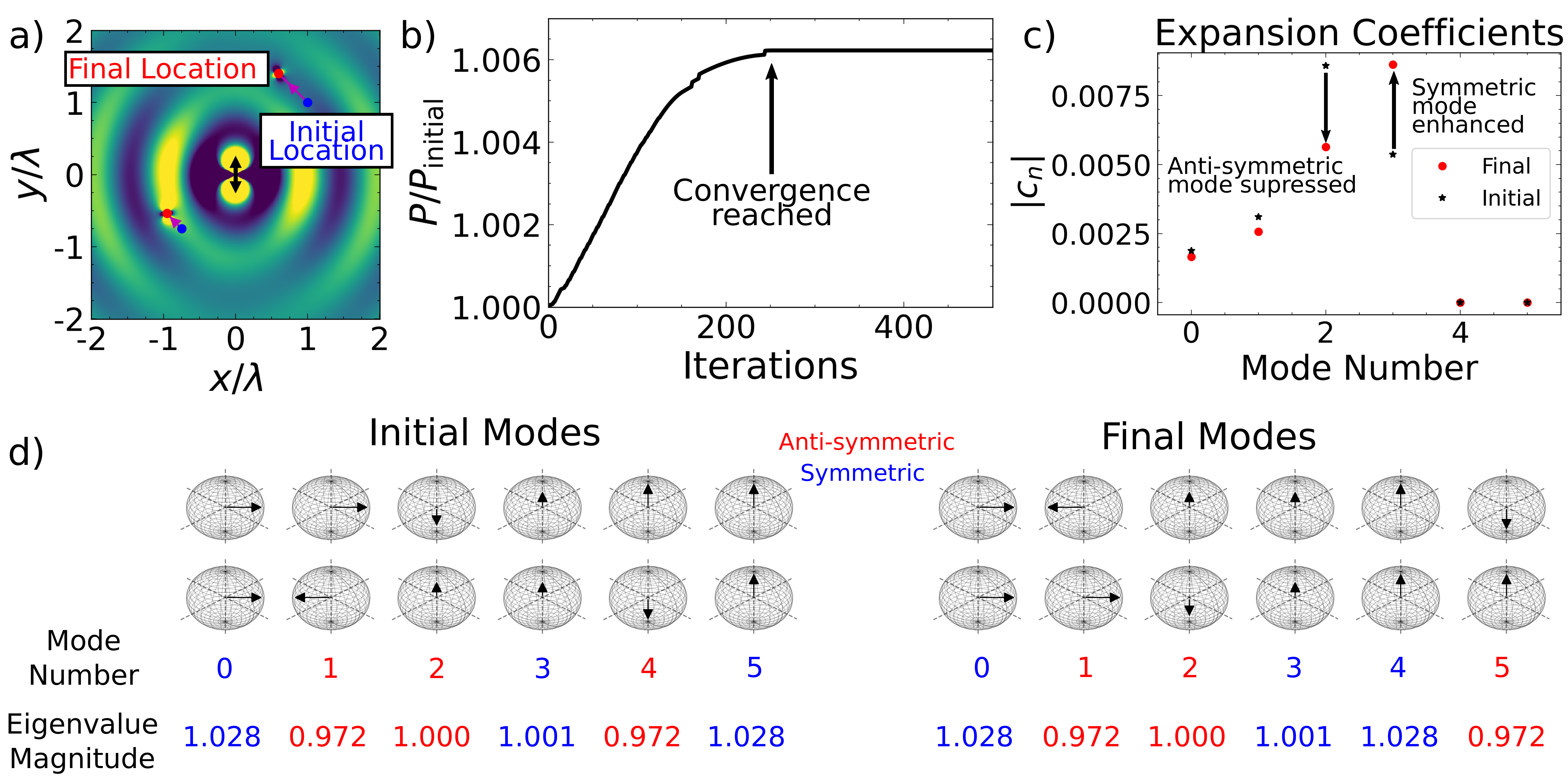}
    \caption{A simple example of the optimisation procedure, and the tools that will be used to characterise more complex structures.  
    The aim is to enhance the power emission of the dipole emitter at the origin (black arrow).
    We consider two scatterers (dots) that support only an electric dipole ($\alpha_H$ = 0). (a) shows the path the scatterers take during the optimisation.  (b) shows the evolution of the power emission.
    (c) Shows how the optimisation process changes the relative dominance of the modes in the response of the system, as characterised by their expansion coefficient.
    (d) Shows how the eigenmodes of the system, with their associated eigenvalues, are modified by the design procedure. Note the two classes of mode: symmetric (i.e. initial mode 0) and anti-symmetric (i.e. initial mode 1).  Eigenmodes are indicated as pairs of spheres, indicating each of the two scatterers with the arrows representing the three components of the electric field vector.}
    \label{fig:twoElectricScatterers}
\end{figure}

Before proceeding to the design of complex structures, we consider the simple case of two electric scatterers.
The eigenvalues for this situation are given by (\ref{eq:evals}), and the eigenvectors can be found from the response matrix in (\ref{eq:R_two_electric}).
The progression of the optimisation procedure is shown in Figure \ref{fig:twoElectricScatterers}(b).
The initial and final eigenmodes and eigenvalues are shown in Figure \ref{fig:twoElectricScatterers}(d-f).
It can be seen that the eigenvalues of the modes change very little as the optimisation process progresses.
This is unsurprising in this case, as from Figure \ref{fig:twoElectricScatterers}(a) it can be seen that the optimisation has resulted in what is almost just a translation of the two scatterers.
Instead of the modes themselves being extensively modified, the change occurs in which modes are excited.
Examining how the expansion coefficients change, we find that the expansion coefficient of mode 2, which is anti-symmetric, is suppressed.
On the other hand, the expansion coefficient of mode 3, which is symmetric, is enhanced.
This is the origin of the power enhancement.
The magnitude of the enhancement is small as the size of the eigenvalue of mode 3 is very close to unity meaning that the collective response of this mode is small, indicating weak coupling between the scatterers.

This simple example demonstrates three important things.
Firstly, there are two mechanisms by which the optimisation procedure may increase the figure of merit.  
The eigenmodes of the system may themselves be changed.
This may present as a change of the spatial distribution of the mode, or the increase of eigenvalues.
Next, with only two scatterers the enhancements we can achieve are small.  
Indeed, this is congruent with the interpretation that large eigenvalues, corresponding to large collective responses, represent multiple scattering events.
A significant number of scatterers are necessary to achieve a large enhancement.
Also, our numerical implementation of the procedure outlined in Figure \ref{fig:setup} makes no attempt to avoid local minima (for example by adding random perturbations to the optimisation process \cite{Pincus1970}).
Indeed, we see that the solution presented in Figure \ref{fig:twoElectricScatterers}(a) is probably a local minima, rather than a global one.

\begin{figure}
    \centering
    \includegraphics[width=\linewidth]{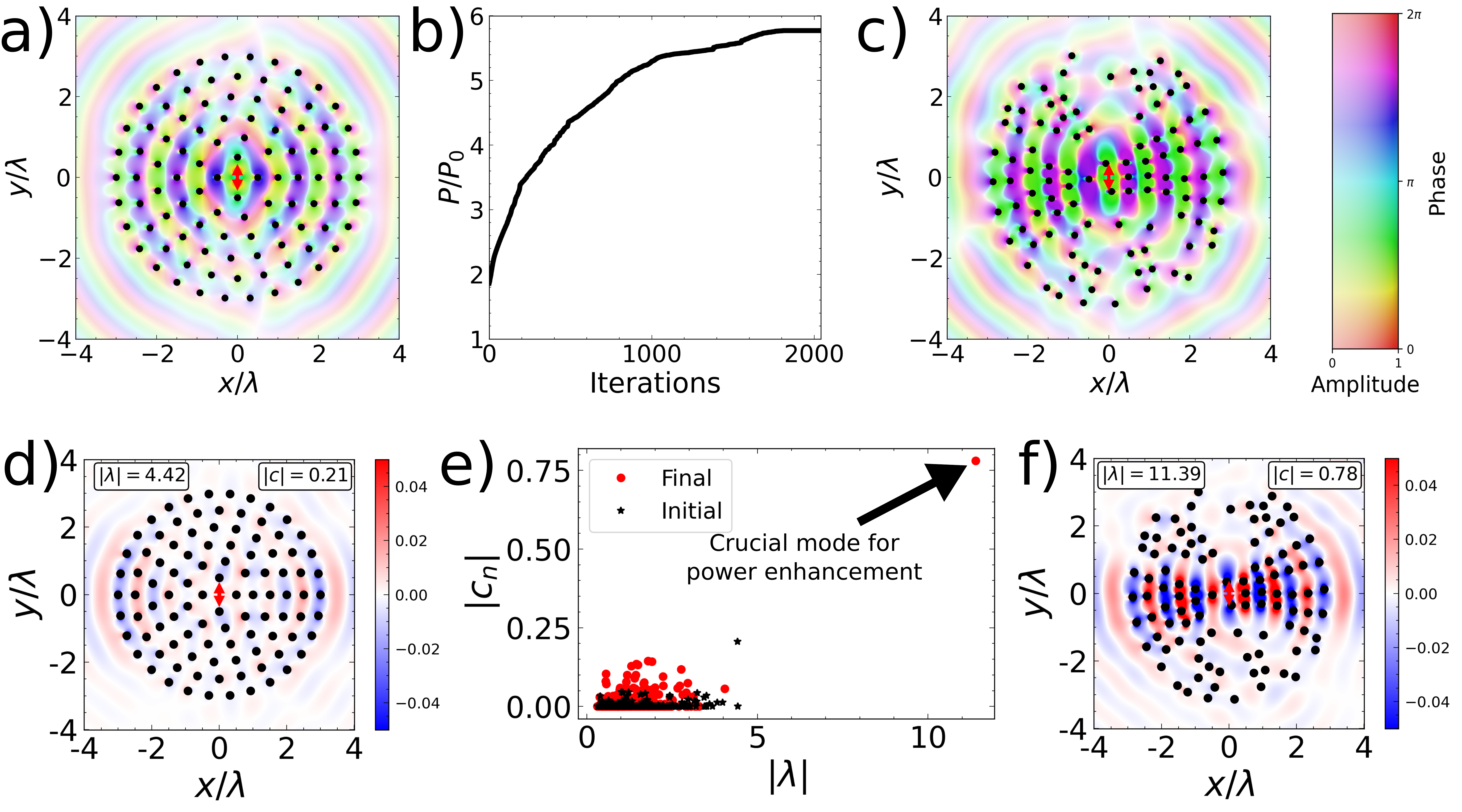}
    \caption{The result of applying our design methodology to enhance power emission of a dipole using 100 scatterers. In all plots, scatterers are shown as black circles and the dipole emitter as a red arrow. (a) Shows the $\boldsymbol{\hat{y}}$ component of the electric field in the initial configuration, (b) shows the progress of the power enhancement as the optimisation progresses and (c) shows ${\rm Re}[\boldsymbol{E} \cdot \boldsymbol{\hat{y}}]$ in the optimised configuration. (d) and (f) show ${\rm Re}[\boldsymbol{E} \cdot \boldsymbol{\hat{y}}]$ of the mode with the largest expansion coefficient in the initial and optimised structure respectively. (e) shows how the eigenvalues and expansion coefficients change due to the optimisation.  The final mode, plotted in (f), with eigenvalue $\sim 10$ and expansion coefficient $\sim 0.75$ is responsible for the power enhancement.}
    \label{fig:pwr}
\end{figure}

To demonstrate the ability of our method to manipulate the near field of a source, we show how structures can be designed to enhance the power emission of a dipole.
The results of this are shown in Figure \ref{fig:pwr}.
An arrangement of 100 dipolar scatterers has been designed using our proposed framework to provide a factor of $\sim 3$ enhancement of the power emission of the dipole emitter.
This factor of enhancement is far smaller than can be achieved with 3D bulk structures \cite{Mignuzzi2019}, but is of the order of similar works that have utilised genetic algorithm techniques \cite{Wiecha2018}.
Examining the change in the eigenmode with the largest expansion coefficient, by comparing Figure \ref{fig:pwr}(d) with Figure \ref{fig:pwr}(f), two key qualitative features of modes that enhance power emission can be determined.
Firstly, the mode has a large eigenvalue.
This corresponds to a large collective response of the scatterers.
Indeed, this effect can be seen in the field shown in Figure \ref{fig:pwr}(f), as the scatterers to the left and right of the dipole emitter are illuminated strongly.
This effect is clearly not present in Figure \ref{fig:pwr}(d).
Secondly, not only does this mode have a large eigenvalue, but it exhibits a strong localisation at the location of the emitter.
Clearly, this mode and the field from the dipole have a large overlap, which is also demonstrated by the large expansion coefficient of the  mode shown in Figure \ref{fig:pwr}(f).
This large collective response of the scatterer system, as well as being strongly excited by the dipole's field, leads to the enhancement of power emission.

\begin{figure}
    \centering
    \includegraphics[width=0.8\linewidth]{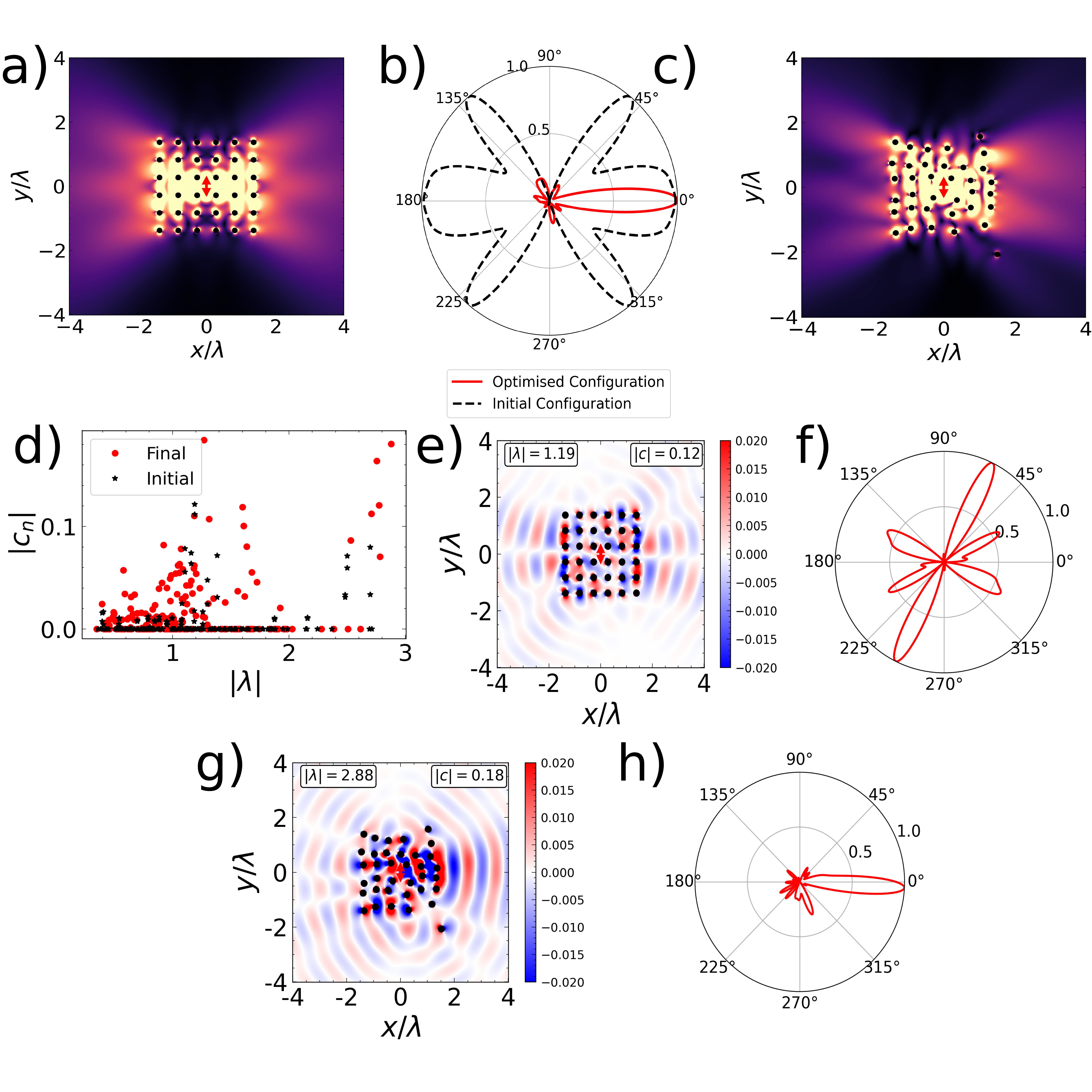}
    \caption{Re-structuring far-field of a dipole emitter to be directed along $\theta = 0^\circ$, using 36 scatterers.  
    (a) shows $|\boldsymbol{E}|$ for the initial configuration, where scatterers are black circles and the emitter is shown as a red arrow. 
    (b) shows a comparison of the far-field distribution of the Poynting vector for the initial configuration (black dashed line) and the optimised configuration (red line).  The width of the beam in the optimised structure is $\sim 24^\circ$. 
    (c) shows $|\boldsymbol{E}|$ in the optimized structure.
    The evolution of the properties of the eigenmode with the largest expansion coefficient is shown in (d-h).  Both the field distribution in the plane of the scatterers and the normalized far-field Poynting vector are shown.
    (e-f) show the mode with the largest expansion coefficient in the initial strucutre, with (g-h) showing the mode with the largest expansion coefficient in the final structure.
    It is clear that the mode shown in (g-h) is responsible for the strong directivity along $\theta = 0^\circ$ in the optimized structure.}
    \label{fig:dir}
\end{figure}

Next, we demonstrate the ability of our method to manipulate the far-field of a dipole emitter.
We apply our method to enhance directivity of a dipole emitter along a given direction.
The results of this optimisation are shown in Figure \ref{fig:dir}.
We have sought to enhance directivity along the $\theta = 0$ direction, where $\theta$ is the polar angle in the $x-y$ plane.
Beginning from a square array, with a far-field radiation pattern shown in Figure \ref{fig:dir}(b), a point in the far-field has been chosen with the power radiated to this location being the figure of merit.
The result is a clear enhancement of the directivity along the $\theta = 0$ direction.
While back-lobes are present, they are $\sim 5$ times smaller than the main lobe.
The beam-width is $\sim \pm 12^\circ$.
The beam width at half power in the $x-y$ plane is $\theta_{\rm HPBW} = 24^\circ$, and in the $x-z$ plane is $\phi_{HPBW} = 1.6^\circ$.
To use the common \cite{antennas} antenna directivity estimate ${D} = 10 \log ( 41000 / (\theta_{\rm HPBW} \phi_{\rm HPBW} D_0) )$, where  $D_0$ is the reference directivity.
For an isotropic source $D_0 = 1$ and for a dipole $D_0 = 6$.
This approximation ignores back-lobes.
Using this, we estimate the structure shown in Figure \ref{fig:dir}(c) has directivity 70 dB above isotropic and 65 dB above a dipole.

In addition to designing a structure with the desired far-field properties, by applying our understanding of the eigenmodes of the system the mechanism for effective performance can be revealed.
The change in the eigenvalues and expansion coefficients of the modes is shown in Figure \ref{fig:dir}(d).
We note that the eigenvalues and expansion coefficients do not change magnitude considerably as a result of the optimisation.
This is perhaps not very surprising, as the aim of the optimisation was not to enhance power emission but rather to re-structure the field.
So, rather than designing modes with a large eigenvalue and therefore a large power emission enhancement due to a collective response, the optimisation procedure has designed modes with a certain shape.
This is demonstrated in Figure \ref{fig:dir}(e-h), where the leading order modes, characterised by expansion coefficient, of the system before and after the optimisation are plotted, with the far-field angular distribution of the associated Poynting vector inset.
Both of the modes have similar expansion coefficients and eigenvalues, but very different spatial distributions.
Indeed, for this application the optimisation procedure re-shapes the modes rather than enhancing multiple scattering effects.

\begin{figure}
    \centering
    \includegraphics[width=\linewidth]{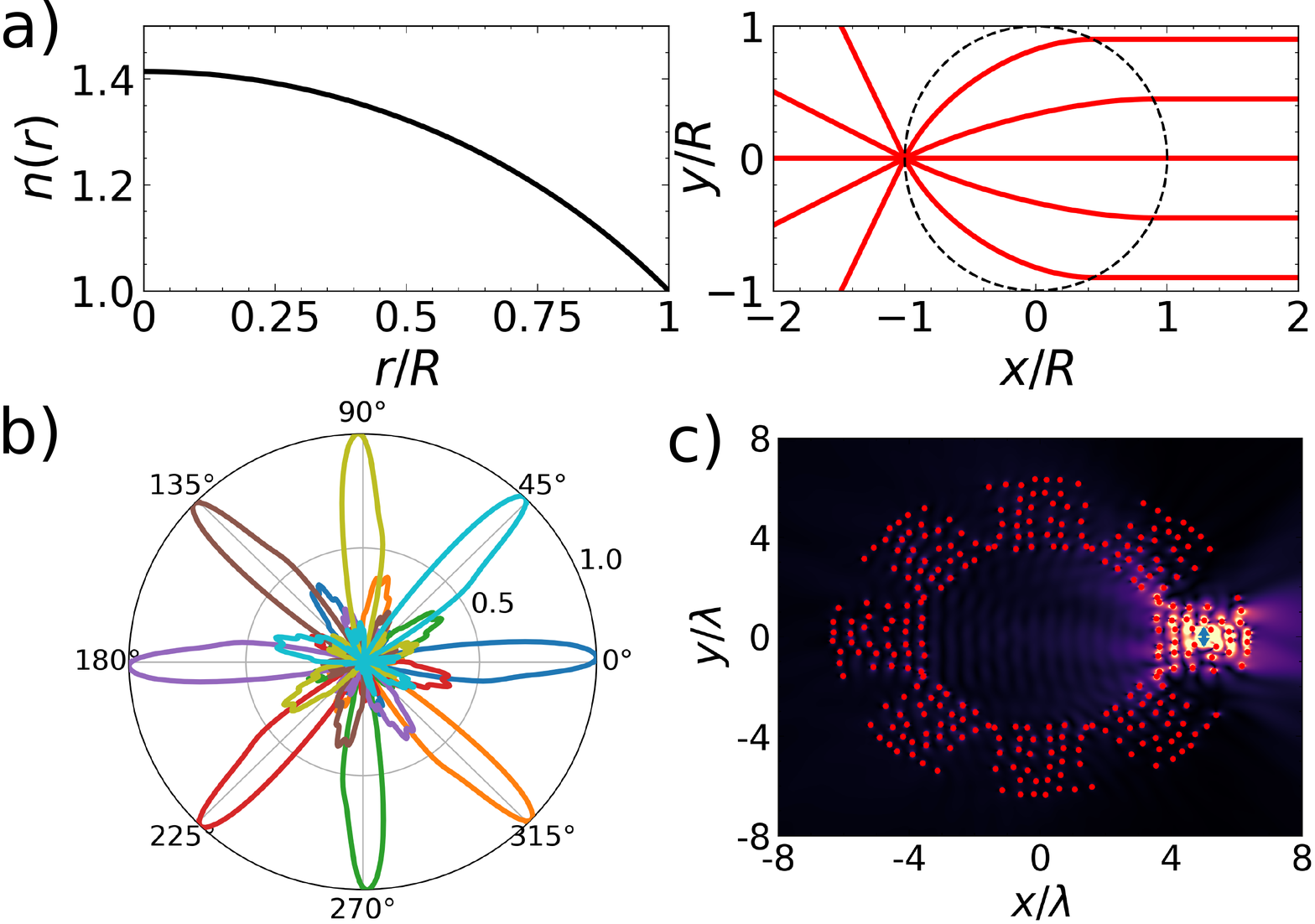}
    \caption{Multiplexing the design shown in Figure \ref{fig:dir} to construct a Luneburg lens. (a) demonstrates the function of a normal Luneburg lens of radius $R$, with a refractive index graded according to the inset equation.  A point source is converted into a beam in a single direction. (b) and (c) The result of multiplexing the structure proposed in Figure \ref{fig:dir} to produce a Luneburg lens with a discrete angular resolution of 45$^\circ$.  By rotating the source and changing it's location inside the array the far-field Poynting vectors indicated in (b) can be observed.}
    \label{fig:luneburg}
\end{figure}

Now, we shall demonstrate how the results of our optimisation procedure may be utilised to achieve more complex functionality.
By rotating the structure designed to enhance directivity, as well as the source, a device can be constructed which has the same functionality as a Luneburg lens \cite{Luneburg1964}.
Using this device, a point source may be converted to a plane wave with a specific propagation direction.
This is shown in Figure \ref{fig:luneburg}.
A conventional Luneburg lens is made by grading the refractive index profile, so that a point source placed upon the surface of the lens is converted into a plane wave.  Its ray diagram , together with the refractive index profile required to achieve this function,  is shown in Figure \ref{fig:luneburg}(a).

Exploiting the spherical symmetry of the refractive index profile, by rotating the source around the edge of the lens the plane wave may also be rotated.
Multiplexing the result shown in Figure \ref{fig:dir} can achieve something qualitatively similar, as demonstrated in Figure \ref{fig:luneburg}(b-c).
One can use the indicated structure to convert a dipole source into a beam directed along a single angle.
By placing the source at different locations within the structure, with the correct orientation, the beam may be rotated.
It should be noted that additional side lobes can be seen in the far-field Poynting vector in Figure \ref{fig:luneburg}(b) compared to Figure \ref{fig:dir}(b).
This is due to interaction between the different multiplexed elements that make up the structure and can be reduced if the spacing between elements was increased.
The structure proposed in Figure \ref{fig:luneburg}(c) can be fabricated without having to grade an index, instead 288 scatterers must be arranged as indicated.
The multiplexed device has a radius comparable to conventional Luneburg lenses, at $\sim 6 \lambda$.
While the fabrication is more straightforward, the angular resolution is not continuous as for the usual Luneburg lens.
Instead, an angular resolution of 45$^\circ$ can be achieved, although by making the structure larger a higher resolution could be achieved.
It was found that the relationship between device radius and angular resolution was well approximated by the following power law: ${\rm resolution} \ ({\rm degrees}) = 83.6 \times (R/\lambda)^{-0.66}$.
Therefore, to achieve a resolution of $2^\circ$, the device would have to be $\sim 50 \lambda$.

\section{Summary and Conclusions}

In this work, we have derived a method of designing metasurfaces comprised of dipolar scatterers.
This has been applied to structure both the near-field and the far field of a dipole emitter.
In the near-field, power emission has been enhanced by a factor of $\sim 3$ and the far-field radiation pattern has been re-structured.
We have also demonstrated that structures designed in this way may be multiplexed to achieve more complex functionality.
As an example, we approximate the functionality of a Luneburg lens using an array of dipole scatterers.

As well as an iterative design methodology, we propose an interesting physical interpretation of the eigenvalues and eigenvectors of the matrix defining the electromagnetic response of the scattering system.
The eigenvalues of the system correspond to eigen-polarizabilities which we attribute to several scatterers responding collectively.
A large collective response corresponds to a large eigenvalue.
By analysing how the eigenvalues and eigenvectors change over the optimisation procedure, we have identified that power emission is enhanced by a large collective response of the scatterers, corresponding to a large eigenvalue while directivity is achieved by modifying the spatial distribution of the modes, without significant change to the eigenvalues.

The applicability of both our design technique and theoretical understanding are not limited to engineering dipole radiation.
A perturbative approach to designing electromagetic field properties might be applied to engineering mode distributions in optical fibers or metalenses.
If more arbitrary field distributions could be successfully designed, then this method might find utility in constructing metasurface holograms or to perform wavefront shaping when imaging through disordered media.
As our approach automatically takes non-locality into account, it may be used to develop and provide insight into non-local metasurfaces. 

\section*{Acknowledgements}

J.R.C would like to thank Ben Pearce and Josh Glasbey for many useful conversations.

We acknowledge financial support from the Engineering and Physical Sciences Research Council (EPSRC) of the United Kingdom, via the EPSRC Centre for Doctoral Training in Metamaterials (Grant No. EP/L015331/1). 
J.R.C also wishes to acknowledge financial support from Defence Science Technology Laboratory (DSTL). 
S.A.R.H acknowledges financial support from the Royal Society (RPG-2016-186).

\bibliography{sorted_refs}

\begin{thebibliography}{56}%
\makeatletter
\providecommand \@ifxundefined [1]{%
 \@ifx{#1\undefined}
}%
\providecommand \@ifnum [1]{%
 \ifnum #1\expandafter \@firstoftwo
 \else \expandafter \@secondoftwo
 \fi
}%
\providecommand \@ifx [1]{%
 \ifx #1\expandafter \@firstoftwo
 \else \expandafter \@secondoftwo
 \fi
}%
\providecommand \natexlab [1]{#1}%
\providecommand \enquote  [1]{``#1''}%
\providecommand \bibnamefont  [1]{#1}%
\providecommand \bibfnamefont [1]{#1}%
\providecommand \citenamefont [1]{#1}%
\providecommand \href@noop [0]{\@secondoftwo}%
\providecommand \href [0]{\begingroup \@sanitize@url \@href}%
\providecommand \@href[1]{\@@startlink{#1}\@@href}%
\providecommand \@@href[1]{\endgroup#1\@@endlink}%
\providecommand \@sanitize@url [0]{\catcode `\\12\catcode `\$12\catcode
  `\&12\catcode `\#12\catcode `\^12\catcode `\_12\catcode `\%12\relax}%
\providecommand \@@startlink[1]{}%
\providecommand \@@endlink[0]{}%
\providecommand \url  [0]{\begingroup\@sanitize@url \@url }%
\providecommand \@url [1]{\endgroup\@href {#1}{\urlprefix }}%
\providecommand \urlprefix  [0]{URL }%
\providecommand \Eprint [0]{\href }%
\providecommand \doibase [0]{http://dx.doi.org/}%
\providecommand \selectlanguage [0]{\@gobble}%
\providecommand \bibinfo  [0]{\@secondoftwo}%
\providecommand \bibfield  [0]{\@secondoftwo}%
\providecommand \translation [1]{[#1]}%
\providecommand \BibitemOpen [0]{}%
\providecommand \bibitemStop [0]{}%
\providecommand \bibitemNoStop [0]{.\EOS\space}%
\providecommand \EOS [0]{\spacefactor3000\relax}%
\providecommand \BibitemShut  [1]{\csname bibitem#1\endcsname}%
\let\auto@bib@innerbib\@empty
\bibitem [{\citenamefont {Quevedo-Teruel}\ \emph {et~al.}(2019)\citenamefont
  {Quevedo-Teruel}, \citenamefont {Chen}, \citenamefont {D{\'{\i}}az-Rubio},
  \citenamefont {Gok}, \citenamefont {Grbic}, \citenamefont {Minatti},
  \citenamefont {Martini}, \citenamefont {Maci}, \citenamefont {Eleftheriades},
  \citenamefont {Chen}, \citenamefont {Zheludev}, \citenamefont {Papasimakis},
  \citenamefont {Choudhury}, \citenamefont {Kudyshev}, \citenamefont {Saha},
  \citenamefont {Reddy}, \citenamefont {Boltasseva}, \citenamefont {Shalaev},
  \citenamefont {Kildishev}, \citenamefont {Sievenpiper}, \citenamefont
  {Caloz}, \citenamefont {Al{\`{u}}}, \citenamefont {He}, \citenamefont {Zhou},
  \citenamefont {Valerio}, \citenamefont {Rajo-Iglesias}, \citenamefont
  {Sipus}, \citenamefont {Mesa}, \citenamefont {Rodr{\'{\i}}guez-Berral},
  \citenamefont {Medina}, \citenamefont {Asadchy}, \citenamefont {Tretyakov},\
  and\ \citenamefont {Craeye}}]{Quevedo-Teruel2019}%
  \BibitemOpen
  \bibfield  {author} {\bibinfo {author} {\bibfnamefont {O.}~\bibnamefont
  {Quevedo-Teruel}}, \bibinfo {author} {\bibfnamefont {H.}~\bibnamefont
  {Chen}}, \bibinfo {author} {\bibfnamefont {A.}~\bibnamefont
  {D{\'{\i}}az-Rubio}}, \bibinfo {author} {\bibfnamefont {G.}~\bibnamefont
  {Gok}}, \bibinfo {author} {\bibfnamefont {A.}~\bibnamefont {Grbic}}, \bibinfo
  {author} {\bibfnamefont {G.}~\bibnamefont {Minatti}}, \bibinfo {author}
  {\bibfnamefont {E.}~\bibnamefont {Martini}}, \bibinfo {author} {\bibfnamefont
  {S.}~\bibnamefont {Maci}}, \bibinfo {author} {\bibfnamefont {G.~V.}\
  \bibnamefont {Eleftheriades}}, \bibinfo {author} {\bibfnamefont
  {M.}~\bibnamefont {Chen}}, \bibinfo {author} {\bibfnamefont {N.~I.}\
  \bibnamefont {Zheludev}}, \bibinfo {author} {\bibfnamefont {N.}~\bibnamefont
  {Papasimakis}}, \bibinfo {author} {\bibfnamefont {S.}~\bibnamefont
  {Choudhury}}, \bibinfo {author} {\bibfnamefont {Z.~A.}\ \bibnamefont
  {Kudyshev}}, \bibinfo {author} {\bibfnamefont {S.}~\bibnamefont {Saha}},
  \bibinfo {author} {\bibfnamefont {H.}~\bibnamefont {Reddy}}, \bibinfo
  {author} {\bibfnamefont {A.}~\bibnamefont {Boltasseva}}, \bibinfo {author}
  {\bibfnamefont {V.~M.}\ \bibnamefont {Shalaev}}, \bibinfo {author}
  {\bibfnamefont {A.~V.}\ \bibnamefont {Kildishev}}, \bibinfo {author}
  {\bibfnamefont {D.}~\bibnamefont {Sievenpiper}}, \bibinfo {author}
  {\bibfnamefont {C.}~\bibnamefont {Caloz}}, \bibinfo {author} {\bibfnamefont
  {A.}~\bibnamefont {Al{\`{u}}}}, \bibinfo {author} {\bibfnamefont
  {Q.}~\bibnamefont {He}}, \bibinfo {author} {\bibfnamefont {L.}~\bibnamefont
  {Zhou}}, \bibinfo {author} {\bibfnamefont {G.}~\bibnamefont {Valerio}},
  \bibinfo {author} {\bibfnamefont {E.}~\bibnamefont {Rajo-Iglesias}}, \bibinfo
  {author} {\bibfnamefont {Z.}~\bibnamefont {Sipus}}, \bibinfo {author}
  {\bibfnamefont {F.}~\bibnamefont {Mesa}}, \bibinfo {author} {\bibfnamefont
  {R.}~\bibnamefont {Rodr{\'{\i}}guez-Berral}}, \bibinfo {author}
  {\bibfnamefont {F.}~\bibnamefont {Medina}}, \bibinfo {author} {\bibfnamefont
  {V.}~\bibnamefont {Asadchy}}, \bibinfo {author} {\bibfnamefont
  {S.}~\bibnamefont {Tretyakov}}, \ and\ \bibinfo {author} {\bibfnamefont
  {C.}~\bibnamefont {Craeye}},\ }\href {\doibase 10.1088/2040-8986/ab161d}
  {\bibfield  {journal} {\bibinfo  {journal} {Journal of Optics}\ }\textbf
  {\bibinfo {volume} {21}},\ \bibinfo {pages} {073002} (\bibinfo {year}
  {2019})}\BibitemShut {NoStop}%
\bibitem [{\citenamefont {Mosk}\ \emph {et~al.}(2012)\citenamefont {Mosk},
  \citenamefont {Lagendijk}, \citenamefont {Lerosey},\ and\ \citenamefont
  {Fink}}]{Mosk2012}%
  \BibitemOpen
  \bibfield  {author} {\bibinfo {author} {\bibfnamefont {A.~P.}\ \bibnamefont
  {Mosk}}, \bibinfo {author} {\bibfnamefont {A.}~\bibnamefont {Lagendijk}},
  \bibinfo {author} {\bibfnamefont {G.}~\bibnamefont {Lerosey}}, \ and\
  \bibinfo {author} {\bibfnamefont {M.}~\bibnamefont {Fink}},\ }\href {\doibase
  10.1038/nphoton.2012.88} {\bibfield  {journal} {\bibinfo  {journal} {Nature
  Photonics}\ }\textbf {\bibinfo {volume} {6}},\ \bibinfo {pages} {283}
  (\bibinfo {year} {2012})}\BibitemShut {NoStop}%
\bibitem [{\citenamefont {Vellekoop}\ and\ \citenamefont
  {Mosk}(2007)}]{Vellekoop2007}%
  \BibitemOpen
  \bibfield  {author} {\bibinfo {author} {\bibfnamefont {I.~M.}\ \bibnamefont
  {Vellekoop}}\ and\ \bibinfo {author} {\bibfnamefont {A.~P.}\ \bibnamefont
  {Mosk}},\ }\href {\doibase 10.1364/ol.32.002309} {\bibfield  {journal}
  {\bibinfo  {journal} {Optics Letters}\ }\textbf {\bibinfo {volume} {32}},\
  \bibinfo {pages} {2309} (\bibinfo {year} {2007})}\BibitemShut {NoStop}%
\bibitem [{\citenamefont {Vellekoop}\ \emph {et~al.}(2008)\citenamefont
  {Vellekoop}, \citenamefont {van Putten}, \citenamefont {Lagendijk},\ and\
  \citenamefont {Mosk}}]{Vellekoop2008}%
  \BibitemOpen
  \bibfield  {author} {\bibinfo {author} {\bibfnamefont {I.~M.}\ \bibnamefont
  {Vellekoop}}, \bibinfo {author} {\bibfnamefont {E.~G.}\ \bibnamefont {van
  Putten}}, \bibinfo {author} {\bibfnamefont {A.}~\bibnamefont {Lagendijk}}, \
  and\ \bibinfo {author} {\bibfnamefont {A.~P.}\ \bibnamefont {Mosk}},\ }\href
  {\doibase 10.1364/oe.16.000067} {\bibfield  {journal} {\bibinfo  {journal}
  {Optics Express}\ }\textbf {\bibinfo {volume} {16}},\ \bibinfo {pages} {67}
  (\bibinfo {year} {2008})}\BibitemShut {NoStop}%
\bibitem [{\citenamefont {Vellekoop}\ and\ \citenamefont
  {Mosk}(2008)}]{Vellekoop2008a}%
  \BibitemOpen
  \bibfield  {author} {\bibinfo {author} {\bibfnamefont {I.~M.}\ \bibnamefont
  {Vellekoop}}\ and\ \bibinfo {author} {\bibfnamefont {A.~P.}\ \bibnamefont
  {Mosk}},\ }\href {\doibase 10.1103/PhysRevLett.101.120601} {\bibfield
  {journal} {\bibinfo  {journal} {Physical Review Letters}\ }\textbf {\bibinfo
  {volume} {101}},\ \bibinfo {pages} {1} (\bibinfo {year} {2008})},\ \Eprint
  {http://arxiv.org/abs/0804.2412} {arXiv:0804.2412} \BibitemShut {NoStop}%
\bibitem [{\citenamefont {Pendry}(2008)}]{Pendry2008}%
  \BibitemOpen
  \bibfield  {author} {\bibinfo {author} {\bibfnamefont {J.}~\bibnamefont
  {Pendry}},\ }\href {\doibase 10.1103/physics.1.20} {\bibfield  {journal}
  {\bibinfo  {journal} {Physics}\ }\textbf {\bibinfo {volume} {1}} (\bibinfo
  {year} {2008}),\ 10.1103/physics.1.20}\BibitemShut {NoStop}%
\bibitem [{\citenamefont {Grbic}\ and\ \citenamefont
  {Eleftheriades}(2004)}]{Grbic2004}%
  \BibitemOpen
  \bibfield  {author} {\bibinfo {author} {\bibfnamefont {A.}~\bibnamefont
  {Grbic}}\ and\ \bibinfo {author} {\bibfnamefont {G.~V.}\ \bibnamefont
  {Eleftheriades}},\ }\href {\doibase 10.1103/physrevlett.92.117403} {\bibfield
   {journal} {\bibinfo  {journal} {Phys. Rev. Lett.}\ }\textbf {\bibinfo
  {volume} {92}} (\bibinfo {year} {2004}),\
  10.1103/physrevlett.92.117403}\BibitemShut {NoStop}%
\bibitem [{\citenamefont {Ma}\ and\ \citenamefont {Sheng}(2016)}]{Ma2016}%
  \BibitemOpen
  \bibfield  {author} {\bibinfo {author} {\bibfnamefont {G.}~\bibnamefont
  {Ma}}\ and\ \bibinfo {author} {\bibfnamefont {P.}~\bibnamefont {Sheng}},\
  }\href {\doibase 10.1126/sciadv.1501595} {\bibfield  {journal} {\bibinfo
  {journal} {Sci. Adv.}\ }\textbf {\bibinfo {volume} {2}},\ \bibinfo {pages}
  {e1501595} (\bibinfo {year} {2016})}\BibitemShut {NoStop}%
\bibitem [{\citenamefont {Chen}\ \emph {et~al.}(2016)\citenamefont {Chen},
  \citenamefont {Taylor},\ and\ \citenamefont {Yu}}]{Chen2016}%
  \BibitemOpen
  \bibfield  {author} {\bibinfo {author} {\bibfnamefont {H.-T.}\ \bibnamefont
  {Chen}}, \bibinfo {author} {\bibfnamefont {A.~J.}\ \bibnamefont {Taylor}}, \
  and\ \bibinfo {author} {\bibfnamefont {N.}~\bibnamefont {Yu}},\ }\href
  {\doibase 10.1088/0034-4885/79/7/076401} {\bibfield  {journal} {\bibinfo
  {journal} {Rep. Prog. Phys.}\ }\textbf {\bibinfo {volume} {79}},\ \bibinfo
  {pages} {076401} (\bibinfo {year} {2016})}\BibitemShut {NoStop}%
\bibitem [{\citenamefont {Marconi}\ and\ \citenamefont
  {Franklin}(1919)}]{patent}%
  \BibitemOpen
  \bibfield  {author} {\bibinfo {author} {\bibfnamefont {G.}~\bibnamefont
  {Marconi}}\ and\ \bibinfo {author} {\bibfnamefont {C.~S.}\ \bibnamefont
  {Franklin}},\ }\href {https://patents.google.com/patent/US1301473A/en}
  {\enquote {\bibinfo {title} {Reflector for use in wireless telegraphy and
  telephony},}\ } (\bibinfo {year} {1919}),\ \bibinfo {note} {patent Number:
  US1301473A}\BibitemShut {NoStop}%
\bibitem [{\citenamefont {Munk}(2000)}]{FrequencySelectiveSurfaces}%
  \BibitemOpen
  \bibfield  {author} {\bibinfo {author} {\bibfnamefont {B.~A.}\ \bibnamefont
  {Munk}},\ }\href@noop {} {\emph {\bibinfo {title} {Frequency Selective
  Surfaces: Theory and Design}}}\ (\bibinfo  {publisher} {John Wiley and
  Sons},\ \bibinfo {address} {New York},\ \bibinfo {year} {2000})\BibitemShut
  {NoStop}%
\bibitem [{\citenamefont {Anwar}\ \emph {et~al.}(2018)\citenamefont {Anwar},
  \citenamefont {Mao},\ and\ \citenamefont {Ning}}]{Anwar2018}%
  \BibitemOpen
  \bibfield  {author} {\bibinfo {author} {\bibfnamefont {R.~S.}\ \bibnamefont
  {Anwar}}, \bibinfo {author} {\bibfnamefont {L.}~\bibnamefont {Mao}}, \ and\
  \bibinfo {author} {\bibfnamefont {H.}~\bibnamefont {Ning}},\ }\href {\doibase
  10.3390/app8091689} {\bibfield  {journal} {\bibinfo  {journal} {Applied
  Sciences}\ }\textbf {\bibinfo {volume} {8}},\ \bibinfo {pages} {1} (\bibinfo
  {year} {2018})}\BibitemShut {NoStop}%
\bibitem [{\citenamefont {Landy}\ \emph {et~al.}(2008)\citenamefont {Landy},
  \citenamefont {Sajuyigbe}, \citenamefont {Mock}, \citenamefont {Smith},\ and\
  \citenamefont {Padilla}}]{Landy2008}%
  \BibitemOpen
  \bibfield  {author} {\bibinfo {author} {\bibfnamefont {N.~I.}\ \bibnamefont
  {Landy}}, \bibinfo {author} {\bibfnamefont {S.}~\bibnamefont {Sajuyigbe}},
  \bibinfo {author} {\bibfnamefont {J.~J.}\ \bibnamefont {Mock}}, \bibinfo
  {author} {\bibfnamefont {D.~R.}\ \bibnamefont {Smith}}, \ and\ \bibinfo
  {author} {\bibfnamefont {W.~J.}\ \bibnamefont {Padilla}},\ }\href {\doibase
  10.1103/physrevlett.100.207402} {\bibfield  {journal} {\bibinfo  {journal}
  {Phys. Rev. Lett.}\ }\textbf {\bibinfo {volume} {100}} (\bibinfo {year}
  {2008}),\ 10.1103/physrevlett.100.207402}\BibitemShut {NoStop}%
\bibitem [{\citenamefont {Pichler}\ \emph {et~al.}(2019)\citenamefont
  {Pichler}, \citenamefont {K{\"{u}}hmayer}, \citenamefont {B{\"{o}}hm},
  \citenamefont {Brandst{\"{o}}tter}, \citenamefont {Ambichl}, \citenamefont
  {Kuhl},\ and\ \citenamefont {Rotter}}]{Pichler2019}%
  \BibitemOpen
  \bibfield  {author} {\bibinfo {author} {\bibfnamefont {K.}~\bibnamefont
  {Pichler}}, \bibinfo {author} {\bibfnamefont {M.}~\bibnamefont
  {K{\"{u}}hmayer}}, \bibinfo {author} {\bibfnamefont {J.}~\bibnamefont
  {B{\"{o}}hm}}, \bibinfo {author} {\bibfnamefont {A.}~\bibnamefont
  {Brandst{\"{o}}tter}}, \bibinfo {author} {\bibfnamefont {P.}~\bibnamefont
  {Ambichl}}, \bibinfo {author} {\bibfnamefont {U.}~\bibnamefont {Kuhl}}, \
  and\ \bibinfo {author} {\bibfnamefont {S.}~\bibnamefont {Rotter}},\ }\href
  {\doibase 10.1038/s41586-019-0971-3} {\bibfield  {journal} {\bibinfo
  {journal} {Nature}\ }\textbf {\bibinfo {volume} {567}},\ \bibinfo {pages}
  {351} (\bibinfo {year} {2019})}\BibitemShut {NoStop}%
\bibitem [{\citenamefont {Yu}\ \emph {et~al.}(2011)\citenamefont {Yu},
  \citenamefont {Genevet}, \citenamefont {Kats}, \citenamefont {Aieta},
  \citenamefont {Tetienne}, \citenamefont {Capasso},\ and\ \citenamefont
  {Gaburro}}]{Yu2011}%
  \BibitemOpen
  \bibfield  {author} {\bibinfo {author} {\bibfnamefont {N.}~\bibnamefont
  {Yu}}, \bibinfo {author} {\bibfnamefont {P.}~\bibnamefont {Genevet}},
  \bibinfo {author} {\bibfnamefont {M.~a.}\ \bibnamefont {Kats}}, \bibinfo
  {author} {\bibfnamefont {F.}~\bibnamefont {Aieta}}, \bibinfo {author}
  {\bibfnamefont {J.-P.}\ \bibnamefont {Tetienne}}, \bibinfo {author}
  {\bibfnamefont {F.}~\bibnamefont {Capasso}}, \ and\ \bibinfo {author}
  {\bibfnamefont {Z.}~\bibnamefont {Gaburro}},\ }\href {\doibase
  10.1126/science.1210713} {\bibfield  {journal} {\bibinfo  {journal}
  {Science}\ }\textbf {\bibinfo {volume} {334}},\ \bibinfo {pages} {333}
  (\bibinfo {year} {2011})}\BibitemShut {NoStop}%
\bibitem [{\citenamefont {Chen}\ \emph {et~al.}(2020)\citenamefont {Chen},
  \citenamefont {Zhu},\ and\ \citenamefont {Capasso}}]{Chen2020}%
  \BibitemOpen
  \bibfield  {author} {\bibinfo {author} {\bibfnamefont {W.~T.}\ \bibnamefont
  {Chen}}, \bibinfo {author} {\bibfnamefont {A.~Y.}\ \bibnamefont {Zhu}}, \
  and\ \bibinfo {author} {\bibfnamefont {F.}~\bibnamefont {Capasso}},\ }\href
  {\doibase 10.1038/s41578-020-0203-3} {\bibfield  {journal} {\bibinfo
  {journal} {Nature Reviews Materials}\ } (\bibinfo {year} {2020}),\
  10.1038/s41578-020-0203-3}\BibitemShut {NoStop}%
\bibitem [{\citenamefont {Lin}\ \emph {et~al.}(2019)\citenamefont {Lin},
  \citenamefont {Liu}, \citenamefont {Pestourie},\ and\ \citenamefont
  {Johnson}}]{Lin2019}%
  \BibitemOpen
  \bibfield  {author} {\bibinfo {author} {\bibfnamefont {Z.}~\bibnamefont
  {Lin}}, \bibinfo {author} {\bibfnamefont {V.}~\bibnamefont {Liu}}, \bibinfo
  {author} {\bibfnamefont {R.}~\bibnamefont {Pestourie}}, \ and\ \bibinfo
  {author} {\bibfnamefont {S.~G.}\ \bibnamefont {Johnson}},\ }\href {\doibase
  10.1364/oe.27.015765} {\bibfield  {journal} {\bibinfo  {journal} {Opt.
  Express}\ }\textbf {\bibinfo {volume} {27}},\ \bibinfo {pages} {15765}
  (\bibinfo {year} {2019})}\BibitemShut {NoStop}%
\bibitem [{\citenamefont {Faenzi}\ \emph {et~al.}(2019)\citenamefont {Faenzi},
  \citenamefont {Minatti}, \citenamefont {Gonz{\'{a}}lez-Ovejero},
  \citenamefont {Caminita}, \citenamefont {Martini}, \citenamefont {{Della
  Giovampaola}},\ and\ \citenamefont {Maci}}]{Faenzi2019}%
  \BibitemOpen
  \bibfield  {author} {\bibinfo {author} {\bibfnamefont {M.}~\bibnamefont
  {Faenzi}}, \bibinfo {author} {\bibfnamefont {G.}~\bibnamefont {Minatti}},
  \bibinfo {author} {\bibfnamefont {D.}~\bibnamefont {Gonz{\'{a}}lez-Ovejero}},
  \bibinfo {author} {\bibfnamefont {F.}~\bibnamefont {Caminita}}, \bibinfo
  {author} {\bibfnamefont {E.}~\bibnamefont {Martini}}, \bibinfo {author}
  {\bibfnamefont {C.}~\bibnamefont {{Della Giovampaola}}}, \ and\ \bibinfo
  {author} {\bibfnamefont {S.}~\bibnamefont {Maci}},\ }\href {\doibase
  10.1038/s41598-019-46522-z} {\bibfield  {journal} {\bibinfo  {journal}
  {Scientific Reports}\ }\textbf {\bibinfo {volume} {9}},\ \bibinfo {pages} {1}
  (\bibinfo {year} {2019})}\BibitemShut {NoStop}%
\bibitem [{\citenamefont {Ni}\ \emph {et~al.}(2013)\citenamefont {Ni},
  \citenamefont {Kildishev},\ and\ \citenamefont {Shalaev}}]{Ni2013}%
  \BibitemOpen
  \bibfield  {author} {\bibinfo {author} {\bibfnamefont {X.}~\bibnamefont
  {Ni}}, \bibinfo {author} {\bibfnamefont {A.~V.}\ \bibnamefont {Kildishev}}, \
  and\ \bibinfo {author} {\bibfnamefont {V.~M.}\ \bibnamefont {Shalaev}},\
  }\href {\doibase 10.1038/ncomms3807} {\bibfield  {journal} {\bibinfo
  {journal} {Nat Comm}\ }\textbf {\bibinfo {volume} {4}} (\bibinfo {year}
  {2013}),\ 10.1038/ncomms3807}\BibitemShut {NoStop}%
\bibitem [{\citenamefont {Jang}\ \emph {et~al.}(2018)\citenamefont {Jang},
  \citenamefont {Horie}, \citenamefont {Shibukawa}, \citenamefont {Brake},
  \citenamefont {Liu}, \citenamefont {Kamali}, \citenamefont {Arbabi},
  \citenamefont {Ruan}, \citenamefont {Faraon},\ and\ \citenamefont
  {Yang}}]{Jang2018}%
  \BibitemOpen
  \bibfield  {author} {\bibinfo {author} {\bibfnamefont {M.}~\bibnamefont
  {Jang}}, \bibinfo {author} {\bibfnamefont {Y.}~\bibnamefont {Horie}},
  \bibinfo {author} {\bibfnamefont {A.}~\bibnamefont {Shibukawa}}, \bibinfo
  {author} {\bibfnamefont {J.}~\bibnamefont {Brake}}, \bibinfo {author}
  {\bibfnamefont {Y.}~\bibnamefont {Liu}}, \bibinfo {author} {\bibfnamefont
  {S.~M.}\ \bibnamefont {Kamali}}, \bibinfo {author} {\bibfnamefont
  {A.}~\bibnamefont {Arbabi}}, \bibinfo {author} {\bibfnamefont
  {H.}~\bibnamefont {Ruan}}, \bibinfo {author} {\bibfnamefont {A.}~\bibnamefont
  {Faraon}}, \ and\ \bibinfo {author} {\bibfnamefont {C.}~\bibnamefont
  {Yang}},\ }\href {\doibase 10.1038/s41566-017-0078-z} {\bibfield  {journal}
  {\bibinfo  {journal} {Nature Photonics}\ }\textbf {\bibinfo {volume} {12}},\
  \bibinfo {pages} {84} (\bibinfo {year} {2018})},\ \Eprint
  {http://arxiv.org/abs/1706.08640} {arXiv:1706.08640} \BibitemShut {NoStop}%
\bibitem [{\citenamefont {Gerchberg}\ and\ \citenamefont
  {Saxton}(1972)}]{Saxton1972}%
  \BibitemOpen
  \bibfield  {author} {\bibinfo {author} {\bibfnamefont {R.~W.}\ \bibnamefont
  {Gerchberg}}\ and\ \bibinfo {author} {\bibfnamefont {W.~O.}\ \bibnamefont
  {Saxton}},\ }\href {\doibase 10.1070/QE2009v039n06ABEH013642} {\bibfield
  {journal} {\bibinfo  {journal} {Optik}\ }\textbf {\bibinfo {volume} {35}},\
  \bibinfo {pages} {237} (\bibinfo {year} {1972})}\BibitemShut {NoStop}%
\bibitem [{\citenamefont {Overvig}\ \emph {et~al.}(2019)\citenamefont
  {Overvig}, \citenamefont {Shrestha}, \citenamefont {Malek}, \citenamefont
  {Lu}, \citenamefont {Stein}, \citenamefont {Zheng},\ and\ \citenamefont
  {Yu}}]{Overvig2019}%
  \BibitemOpen
  \bibfield  {author} {\bibinfo {author} {\bibfnamefont {A.~C.}\ \bibnamefont
  {Overvig}}, \bibinfo {author} {\bibfnamefont {S.}~\bibnamefont {Shrestha}},
  \bibinfo {author} {\bibfnamefont {S.~C.}\ \bibnamefont {Malek}}, \bibinfo
  {author} {\bibfnamefont {M.}~\bibnamefont {Lu}}, \bibinfo {author}
  {\bibfnamefont {A.}~\bibnamefont {Stein}}, \bibinfo {author} {\bibfnamefont
  {C.}~\bibnamefont {Zheng}}, \ and\ \bibinfo {author} {\bibfnamefont
  {N.}~\bibnamefont {Yu}},\ }\href {\doibase 10.1038/s41377-019-0201-7}
  {\bibfield  {journal} {\bibinfo  {journal} {Light: Science and Applications}\
  }\textbf {\bibinfo {volume} {8}} (\bibinfo {year} {2019}),\
  10.1038/s41377-019-0201-7}\BibitemShut {NoStop}%
\bibitem [{\citenamefont {D{\'{\i}}az-Rubio}\ \emph {et~al.}(2017)\citenamefont
  {D{\'{\i}}az-Rubio}, \citenamefont {Asadchy}, \citenamefont {Elsakka},\ and\
  \citenamefont {Tretyakov}}]{DazRubio2017}%
  \BibitemOpen
  \bibfield  {author} {\bibinfo {author} {\bibfnamefont {A.}~\bibnamefont
  {D{\'{\i}}az-Rubio}}, \bibinfo {author} {\bibfnamefont {V.~S.}\ \bibnamefont
  {Asadchy}}, \bibinfo {author} {\bibfnamefont {A.}~\bibnamefont {Elsakka}}, \
  and\ \bibinfo {author} {\bibfnamefont {S.~A.}\ \bibnamefont {Tretyakov}},\
  }\href {\doibase 10.1126/sciadv.1602714} {\bibfield  {journal} {\bibinfo
  {journal} {Sci. Adv.}\ }\textbf {\bibinfo {volume} {3}},\ \bibinfo {pages}
  {e1602714} (\bibinfo {year} {2017})}\BibitemShut {NoStop}%
\bibitem [{\citenamefont {D{\'{\i}}az-Rubio}\ \emph {et~al.}(2019)\citenamefont
  {D{\'{\i}}az-Rubio}, \citenamefont {Li}, \citenamefont {Shen}, \citenamefont
  {Cummer},\ and\ \citenamefont {Tretyakov}}]{DazRubio2019}%
  \BibitemOpen
  \bibfield  {author} {\bibinfo {author} {\bibfnamefont {A.}~\bibnamefont
  {D{\'{\i}}az-Rubio}}, \bibinfo {author} {\bibfnamefont {J.}~\bibnamefont
  {Li}}, \bibinfo {author} {\bibfnamefont {C.}~\bibnamefont {Shen}}, \bibinfo
  {author} {\bibfnamefont {S.~A.}\ \bibnamefont {Cummer}}, \ and\ \bibinfo
  {author} {\bibfnamefont {S.~A.}\ \bibnamefont {Tretyakov}},\ }\href {\doibase
  10.1126/sciadv.aau7288} {\bibfield  {journal} {\bibinfo  {journal} {Sci.
  Adv.}\ }\textbf {\bibinfo {volume} {5}},\ \bibinfo {pages} {eaau7288}
  (\bibinfo {year} {2019})}\BibitemShut {NoStop}%
\bibitem [{\citenamefont {Molesky}\ \emph {et~al.}(2018)\citenamefont
  {Molesky}, \citenamefont {Lin}, \citenamefont {Piggott}, \citenamefont {Jin},
  \citenamefont {Vuckovi{\'{c}}},\ and\ \citenamefont
  {Rodriguez}}]{Molesky2018}%
  \BibitemOpen
  \bibfield  {author} {\bibinfo {author} {\bibfnamefont {S.}~\bibnamefont
  {Molesky}}, \bibinfo {author} {\bibfnamefont {Z.}~\bibnamefont {Lin}},
  \bibinfo {author} {\bibfnamefont {A.~Y.}\ \bibnamefont {Piggott}}, \bibinfo
  {author} {\bibfnamefont {W.}~\bibnamefont {Jin}}, \bibinfo {author}
  {\bibfnamefont {J.}~\bibnamefont {Vuckovi{\'{c}}}}, \ and\ \bibinfo {author}
  {\bibfnamefont {A.~W.}\ \bibnamefont {Rodriguez}},\ }\href@noop {} {\bibfield
   {journal} {\bibinfo  {journal} {Nature Photonics}\ }\textbf {\bibinfo
  {volume} {12}},\ \bibinfo {pages} {659} (\bibinfo {year} {2018})}\BibitemShut
  {NoStop}%
\bibitem [{\citenamefont {Giles}\ and\ \citenamefont
  {Pierce}(2000)}]{Giles2000}%
  \BibitemOpen
  \bibfield  {author} {\bibinfo {author} {\bibfnamefont {M.~B.}\ \bibnamefont
  {Giles}}\ and\ \bibinfo {author} {\bibfnamefont {N.~A.}\ \bibnamefont
  {Pierce}},\ }\href@noop {} {\bibfield  {journal} {\bibinfo  {journal} {Flow,
  Turbulence and Combustion}\ }\textbf {\bibinfo {volume} {65}},\ \bibinfo
  {pages} {393} (\bibinfo {year} {2000})}\BibitemShut {NoStop}%
\bibitem [{\citenamefont {Lalau-Keraly}\ \emph {et~al.}(2013)\citenamefont
  {Lalau-Keraly}, \citenamefont {Bhargava}, \citenamefont {Miller},\ and\
  \citenamefont {Yablonovitch}}]{Lalau-Keraly2013}%
  \BibitemOpen
  \bibfield  {author} {\bibinfo {author} {\bibfnamefont {C.~M.}\ \bibnamefont
  {Lalau-Keraly}}, \bibinfo {author} {\bibfnamefont {S.}~\bibnamefont
  {Bhargava}}, \bibinfo {author} {\bibfnamefont {O.~D.}\ \bibnamefont
  {Miller}}, \ and\ \bibinfo {author} {\bibfnamefont {E.}~\bibnamefont
  {Yablonovitch}},\ }\href {\doibase 10.1364/oe.21.021693} {\bibfield
  {journal} {\bibinfo  {journal} {Optics Express}\ }\textbf {\bibinfo {volume}
  {21}},\ \bibinfo {pages} {21693} (\bibinfo {year} {2013})}\BibitemShut
  {NoStop}%
\bibitem [{\citenamefont {Mignuzzi}\ \emph {et~al.}(2019)\citenamefont
  {Mignuzzi}, \citenamefont {Vezzoli}, \citenamefont {Horsley}, \citenamefont
  {Barnes}, \citenamefont {Maier},\ and\ \citenamefont
  {Sapienza}}]{Mignuzzi2019}%
  \BibitemOpen
  \bibfield  {author} {\bibinfo {author} {\bibfnamefont {S.}~\bibnamefont
  {Mignuzzi}}, \bibinfo {author} {\bibfnamefont {S.}~\bibnamefont {Vezzoli}},
  \bibinfo {author} {\bibfnamefont {S.~A.}\ \bibnamefont {Horsley}}, \bibinfo
  {author} {\bibfnamefont {W.~L.}\ \bibnamefont {Barnes}}, \bibinfo {author}
  {\bibfnamefont {S.~A.}\ \bibnamefont {Maier}}, \ and\ \bibinfo {author}
  {\bibfnamefont {R.}~\bibnamefont {Sapienza}},\ }\href@noop {} {\bibfield
  {journal} {\bibinfo  {journal} {Nano Letters}\ }\textbf {\bibinfo {volume}
  {19}},\ \bibinfo {pages} {1613} (\bibinfo {year} {2019})}\BibitemShut
  {NoStop}%
\bibitem [{\citenamefont {Landau}\ and\ \citenamefont {Lifshitz}(1994)}]{LL2}%
  \BibitemOpen
  \bibfield  {author} {\bibinfo {author} {\bibfnamefont {L.~D.}\ \bibnamefont
  {Landau}}\ and\ \bibinfo {author} {\bibfnamefont {E.~M.}\ \bibnamefont
  {Lifshitz}},\ }\href@noop {} {\emph {\bibinfo {title} {The Classical Theory
  of Fields}}}\ (\bibinfo  {publisher} {Pergamon Press},\ \bibinfo {address}
  {New York},\ \bibinfo {year} {1994})\BibitemShut {NoStop}%
\bibitem [{\citenamefont {Christiansen}\ \emph {et~al.}(2020)\citenamefont
  {Christiansen}, \citenamefont {Lin}, \citenamefont {Carmes}, \citenamefont
  {Salamin}, \citenamefont {Kooi}, \citenamefont {Joannopoulos}, \citenamefont
  {Solja{\v{c}}i{\'{c}}},\ and\ \citenamefont {Johnson}}]{Christiansen2020}%
  \BibitemOpen
  \bibfield  {author} {\bibinfo {author} {\bibfnamefont {R.~E.}\ \bibnamefont
  {Christiansen}}, \bibinfo {author} {\bibfnamefont {Z.}~\bibnamefont {Lin}},
  \bibinfo {author} {\bibfnamefont {C.~R.}\ \bibnamefont {Carmes}}, \bibinfo
  {author} {\bibfnamefont {Y.}~\bibnamefont {Salamin}}, \bibinfo {author}
  {\bibfnamefont {S.~E.}\ \bibnamefont {Kooi}}, \bibinfo {author}
  {\bibfnamefont {J.~D.}\ \bibnamefont {Joannopoulos}}, \bibinfo {author}
  {\bibfnamefont {M.}~\bibnamefont {Solja{\v{c}}i{\'{c}}}}, \ and\ \bibinfo
  {author} {\bibfnamefont {S.~G.}\ \bibnamefont {Johnson}},\ }\href
  {http://arxiv.org/abs/2007.11661} {\ ,\ \bibinfo {pages} {1} (\bibinfo {year}
  {2020})},\ \Eprint {http://arxiv.org/abs/2007.11661} {arXiv:2007.11661}
  \BibitemShut {NoStop}%
\bibitem [{\citenamefont {Pestourie}\ \emph {et~al.}(2018)\citenamefont
  {Pestourie}, \citenamefont {Pérez-Arancibia}, \citenamefont {Lin},
  \citenamefont {Shin}, \citenamefont {Capasso},\ and\ \citenamefont
  {Johnson}}]{Estourie2018}%
  \BibitemOpen
  \bibfield  {author} {\bibinfo {author} {\bibfnamefont {R.}~\bibnamefont
  {Pestourie}}, \bibinfo {author} {\bibfnamefont {C.}~\bibnamefont
  {Pérez-Arancibia}}, \bibinfo {author} {\bibfnamefont {Z.}~\bibnamefont
  {Lin}}, \bibinfo {author} {\bibfnamefont {W.}~\bibnamefont {Shin}}, \bibinfo
  {author} {\bibfnamefont {F.}~\bibnamefont {Capasso}}, \ and\ \bibinfo
  {author} {\bibfnamefont {S.~G.}\ \bibnamefont {Johnson}},\ }\href@noop {}
  {\bibfield  {journal} {\bibinfo  {journal} {Optics Express}\ }\textbf
  {\bibinfo {volume} {26}},\ \bibinfo {pages} {33732} (\bibinfo {year}
  {2018})}\BibitemShut {NoStop}%
\bibitem [{\citenamefont {Xu}\ \emph {et~al.}(2020)\citenamefont {Xu},
  \citenamefont {Rahmani}, \citenamefont {Ma}, \citenamefont {Smirnova},
  \citenamefont {Kamali}, \citenamefont {Deng}, \citenamefont {Chiang},
  \citenamefont {Huang}, \citenamefont {Zhang}, \citenamefont {Gould},
  \citenamefont {Neshev},\ and\ \citenamefont {Miroshnichenko}}]{Xu}%
  \BibitemOpen
  \bibfield  {author} {\bibinfo {author} {\bibfnamefont {L.}~\bibnamefont
  {Xu}}, \bibinfo {author} {\bibfnamefont {M.}~\bibnamefont {Rahmani}},
  \bibinfo {author} {\bibfnamefont {Y.}~\bibnamefont {Ma}}, \bibinfo {author}
  {\bibfnamefont {D.~A.}\ \bibnamefont {Smirnova}}, \bibinfo {author}
  {\bibfnamefont {K.~Z.}\ \bibnamefont {Kamali}}, \bibinfo {author}
  {\bibfnamefont {F.}~\bibnamefont {Deng}}, \bibinfo {author} {\bibfnamefont
  {Y.~K.}\ \bibnamefont {Chiang}}, \bibinfo {author} {\bibfnamefont
  {L.}~\bibnamefont {Huang}}, \bibinfo {author} {\bibfnamefont
  {H.}~\bibnamefont {Zhang}}, \bibinfo {author} {\bibfnamefont
  {S.}~\bibnamefont {Gould}}, \bibinfo {author} {\bibfnamefont {D.~N.}\
  \bibnamefont {Neshev}}, \ and\ \bibinfo {author} {\bibfnamefont {A.~E.}\
  \bibnamefont {Miroshnichenko}},\ }\href@noop {} {\bibfield  {journal}
  {\bibinfo  {journal} {Advanced Photonics}\ }\textbf {\bibinfo {volume} {2}},\
  \bibinfo {pages} {026003} (\bibinfo {year} {2020})}\BibitemShut {NoStop}%
\bibitem [{\citenamefont {Wiecha}\ \emph {et~al.}(2018)\citenamefont {Wiecha},
  \citenamefont {Arbouet}, \citenamefont {Cuche}, \citenamefont {Paillard},\
  and\ \citenamefont {Girard}}]{Wiecha2018}%
  \BibitemOpen
  \bibfield  {author} {\bibinfo {author} {\bibfnamefont {P.~R.}\ \bibnamefont
  {Wiecha}}, \bibinfo {author} {\bibfnamefont {A.}~\bibnamefont {Arbouet}},
  \bibinfo {author} {\bibfnamefont {A.}~\bibnamefont {Cuche}}, \bibinfo
  {author} {\bibfnamefont {V.}~\bibnamefont {Paillard}}, \ and\ \bibinfo
  {author} {\bibfnamefont {C.}~\bibnamefont {Girard}},\ }\href@noop {}
  {\bibfield  {journal} {\bibinfo  {journal} {Physical Review B}\ }\textbf
  {\bibinfo {volume} {97}},\ \bibinfo {pages} {1} (\bibinfo {year}
  {2018})}\BibitemShut {NoStop}%
\bibitem [{\citenamefont {Gerke}\ and\ \citenamefont
  {Piestun}(2010)}]{Gerke2010a}%
  \BibitemOpen
  \bibfield  {author} {\bibinfo {author} {\bibfnamefont {T.~D.}\ \bibnamefont
  {Gerke}}\ and\ \bibinfo {author} {\bibfnamefont {R.}~\bibnamefont
  {Piestun}},\ }\href {\doibase 10.1038/nphoton.2009.290} {\bibfield  {journal}
  {\bibinfo  {journal} {Nature Photonics}\ }\textbf {\bibinfo {volume} {4}},\
  \bibinfo {pages} {188} (\bibinfo {year} {2010})}\BibitemShut {NoStop}%
\bibitem [{\citenamefont {Born}\ and\ \citenamefont {Wolf}(1999)}]{Born1999}%
  \BibitemOpen
  \bibfield  {author} {\bibinfo {author} {\bibfnamefont {M.}~\bibnamefont
  {Born}}\ and\ \bibinfo {author} {\bibfnamefont {E.}~\bibnamefont {Wolf}},\
  }\href@noop {} {\emph {\bibinfo {title} {{Principles of Optics}}}},\ \bibinfo
  {edition} {7th}\ ed.\ (\bibinfo  {publisher} {Cambridge University Press},\
  \bibinfo {address} {Cambridge},\ \bibinfo {year} {1999})\BibitemShut
  {NoStop}%
\bibitem [{\citenamefont {Mie}(1908)}]{Mie}%
  \BibitemOpen
  \bibfield  {author} {\bibinfo {author} {\bibfnamefont {G.}~\bibnamefont
  {Mie}},\ }\href@noop {} {\bibfield  {journal} {\bibinfo  {journal} {Annalen
  der Physik}\ }\textbf {\bibinfo {volume} {330}},\ \bibinfo {pages} {377}
  (\bibinfo {year} {1908})}\BibitemShut {NoStop}%
\bibitem [{\citenamefont {Bohn}\ \emph {et~al.}(2018)\citenamefont {Bohn},
  \citenamefont {Bucher}, \citenamefont {Chong}, \citenamefont {Komar},
  \citenamefont {Choi}, \citenamefont {Neshev}, \citenamefont {Kivshar},
  \citenamefont {Pertsch},\ and\ \citenamefont {Staude}}]{Bohn2018}%
  \BibitemOpen
  \bibfield  {author} {\bibinfo {author} {\bibfnamefont {J.}~\bibnamefont
  {Bohn}}, \bibinfo {author} {\bibfnamefont {T.}~\bibnamefont {Bucher}},
  \bibinfo {author} {\bibfnamefont {K.~E.}\ \bibnamefont {Chong}}, \bibinfo
  {author} {\bibfnamefont {A.}~\bibnamefont {Komar}}, \bibinfo {author}
  {\bibfnamefont {D.~Y.}\ \bibnamefont {Choi}}, \bibinfo {author}
  {\bibfnamefont {D.~N.}\ \bibnamefont {Neshev}}, \bibinfo {author}
  {\bibfnamefont {Y.~S.}\ \bibnamefont {Kivshar}}, \bibinfo {author}
  {\bibfnamefont {T.}~\bibnamefont {Pertsch}}, \ and\ \bibinfo {author}
  {\bibfnamefont {I.}~\bibnamefont {Staude}},\ }\href@noop {} {\bibfield
  {journal} {\bibinfo  {journal} {Nano Letters}\ }\textbf {\bibinfo {volume}
  {18}},\ \bibinfo {pages} {3461} (\bibinfo {year} {2018})}\BibitemShut
  {NoStop}%
\bibitem [{\citenamefont {Vaskin}\ \emph {et~al.}(2018)\citenamefont {Vaskin},
  \citenamefont {Bohn}, \citenamefont {Chong}, \citenamefont {Bucher},
  \citenamefont {Zilk}, \citenamefont {Choi}, \citenamefont {Neshev},
  \citenamefont {Kivshar}, \citenamefont {Pertsch},\ and\ \citenamefont
  {Staude}}]{Vaskin2018}%
  \BibitemOpen
  \bibfield  {author} {\bibinfo {author} {\bibfnamefont {A.}~\bibnamefont
  {Vaskin}}, \bibinfo {author} {\bibfnamefont {J.}~\bibnamefont {Bohn}},
  \bibinfo {author} {\bibfnamefont {K.~E.}\ \bibnamefont {Chong}}, \bibinfo
  {author} {\bibfnamefont {T.}~\bibnamefont {Bucher}}, \bibinfo {author}
  {\bibfnamefont {M.}~\bibnamefont {Zilk}}, \bibinfo {author} {\bibfnamefont
  {D.~Y.}\ \bibnamefont {Choi}}, \bibinfo {author} {\bibfnamefont {D.~N.}\
  \bibnamefont {Neshev}}, \bibinfo {author} {\bibfnamefont {Y.~S.}\
  \bibnamefont {Kivshar}}, \bibinfo {author} {\bibfnamefont {T.}~\bibnamefont
  {Pertsch}}, \ and\ \bibinfo {author} {\bibfnamefont {I.}~\bibnamefont
  {Staude}},\ }\href@noop {} {\bibfield  {journal} {\bibinfo  {journal} {ACS
  Photonics}\ }\textbf {\bibinfo {volume} {5}},\ \bibinfo {pages} {1359}
  (\bibinfo {year} {2018})}\BibitemShut {NoStop}%
\bibitem [{\citenamefont {Savelev}\ \emph {et~al.}(2014)\citenamefont
  {Savelev}, \citenamefont {Slobozhanyuk}, \citenamefont {Miroshnichenko},
  \citenamefont {Kivshar},\ and\ \citenamefont {Belov}}]{Savelev2014}%
  \BibitemOpen
  \bibfield  {author} {\bibinfo {author} {\bibfnamefont {R.~S.}\ \bibnamefont
  {Savelev}}, \bibinfo {author} {\bibfnamefont {A.~P.}\ \bibnamefont
  {Slobozhanyuk}}, \bibinfo {author} {\bibfnamefont {A.~E.}\ \bibnamefont
  {Miroshnichenko}}, \bibinfo {author} {\bibfnamefont {Y.~S.}\ \bibnamefont
  {Kivshar}}, \ and\ \bibinfo {author} {\bibfnamefont {P.~A.}\ \bibnamefont
  {Belov}},\ }\href {\doibase 10.1103/physrevb.89.035435} {\bibfield  {journal}
  {\bibinfo  {journal} {Phys. Rev. B}\ }\textbf {\bibinfo {volume} {89}}
  (\bibinfo {year} {2014}),\ 10.1103/physrevb.89.035435}\BibitemShut {NoStop}%
\bibitem [{\citenamefont {Belov}\ \emph {et~al.}(2003)\citenamefont {Belov},
  \citenamefont {Maslovski}, \citenamefont {Simovski},\ and\ \citenamefont
  {Tretyakov}}]{Belov2003}%
  \BibitemOpen
  \bibfield  {author} {\bibinfo {author} {\bibfnamefont {P.~A.}\ \bibnamefont
  {Belov}}, \bibinfo {author} {\bibfnamefont {S.~I.}\ \bibnamefont
  {Maslovski}}, \bibinfo {author} {\bibfnamefont {K.~R.}\ \bibnamefont
  {Simovski}}, \ and\ \bibinfo {author} {\bibfnamefont {S.~A.}\ \bibnamefont
  {Tretyakov}},\ }\href {\doibase 10.1134/1.1615545} {\bibfield  {journal}
  {\bibinfo  {journal} {Technical Physics Letters}\ }\textbf {\bibinfo {volume}
  {29}},\ \bibinfo {pages} {718} (\bibinfo {year} {2003})}\BibitemShut
  {NoStop}%
\bibitem [{\citenamefont {Landau}\ and\ \citenamefont {Lifshitz}(1970)}]{LL5}%
  \BibitemOpen
  \bibfield  {author} {\bibinfo {author} {\bibfnamefont {L.}~\bibnamefont
  {Landau}}\ and\ \bibinfo {author} {\bibfnamefont {E.}~\bibnamefont
  {Lifshitz}},\ }\href@noop {} {\emph {\bibinfo {title} {Statistical
  Physics}}}\ (\bibinfo  {publisher} {Pergamon Press},\ \bibinfo {address}
  {Oxford},\ \bibinfo {year} {1970})\BibitemShut {NoStop}%
\bibitem [{\citenamefont {Landau}\ and\ \citenamefont {Lifshitz}(2008)}]{LL8}%
  \BibitemOpen
  \bibfield  {author} {\bibinfo {author} {\bibfnamefont {L.}~\bibnamefont
  {Landau}}\ and\ \bibinfo {author} {\bibfnamefont {E.}~\bibnamefont
  {Lifshitz}},\ }\href@noop {} {\emph {\bibinfo {title} {Electrodynamics of
  Continuous Media}}}\ (\bibinfo  {publisher} {Butterworth Heineman},\ \bibinfo
  {address} {Oxford},\ \bibinfo {year} {2008})\BibitemShut {NoStop}%
\bibitem [{\citenamefont {Tai}(1993)}]{Tai1993}%
  \BibitemOpen
  \bibfield  {author} {\bibinfo {author} {\bibfnamefont {C.-T.}\ \bibnamefont
  {Tai}},\ }\href@noop {} {\emph {\bibinfo {title} {Dyadic Greens Functions in
  Electromagnetic Theory}}}\ (\bibinfo  {publisher} {IEEE Press},\ \bibinfo
  {address} {New York},\ \bibinfo {year} {1993})\BibitemShut {NoStop}%
\bibitem [{\citenamefont {Novotny}\ and\ \citenamefont
  {Hecht}(2006)}]{NanoOps}%
  \BibitemOpen
  \bibfield  {author} {\bibinfo {author} {\bibfnamefont {L.}~\bibnamefont
  {Novotny}}\ and\ \bibinfo {author} {\bibfnamefont {B.}~\bibnamefont
  {Hecht}},\ }\href@noop {} {\emph {\bibinfo {title} {{Principles of
  Nano-optics}}}}\ (\bibinfo  {publisher} {Cambridge University Press},\
  \bibinfo {address} {Cambridge},\ \bibinfo {year} {2006})\BibitemShut
  {NoStop}%
\bibitem [{\citenamefont {Sersic}\ \emph {et~al.}(2011)\citenamefont {Sersic},
  \citenamefont {Tuambilangana}, \citenamefont {Kampfrath},\ and\ \citenamefont
  {Koenderink}}]{Sersic2011}%
  \BibitemOpen
  \bibfield  {author} {\bibinfo {author} {\bibfnamefont {I.}~\bibnamefont
  {Sersic}}, \bibinfo {author} {\bibfnamefont {C.}~\bibnamefont
  {Tuambilangana}}, \bibinfo {author} {\bibfnamefont {T.}~\bibnamefont
  {Kampfrath}}, \ and\ \bibinfo {author} {\bibfnamefont {A.~F.}\ \bibnamefont
  {Koenderink}},\ }\href@noop {} {\bibfield  {journal} {\bibinfo  {journal}
  {Physical Review B}\ }\textbf {\bibinfo {volume} {83}},\ \bibinfo {pages} {1}
  (\bibinfo {year} {2011})}\BibitemShut {NoStop}%
\bibitem [{\citenamefont {Press}\ \emph {et~al.}(2007)\citenamefont {Press},
  \citenamefont {Vetterling}, \citenamefont {Teukolsky},\ and\ \citenamefont
  {Flannery}}]{NumericalRecipes}%
  \BibitemOpen
  \bibfield  {author} {\bibinfo {author} {\bibfnamefont {W.~H.}\ \bibnamefont
  {Press}}, \bibinfo {author} {\bibfnamefont {W.~T.}\ \bibnamefont
  {Vetterling}}, \bibinfo {author} {\bibfnamefont {S.~A.}\ \bibnamefont
  {Teukolsky}}, \ and\ \bibinfo {author} {\bibfnamefont {B.~P.}\ \bibnamefont
  {Flannery}},\ }\href@noop {} {\emph {\bibinfo {title} {Numerical Recipes in
  C}}}\ (\bibinfo  {publisher} {Cambridge University Press},\ \bibinfo
  {address} {Cambridge},\ \bibinfo {year} {2007})\BibitemShut {NoStop}%
\bibitem [{\citenamefont {Barnes}\ \emph {et~al.}(2020)\citenamefont {Barnes},
  \citenamefont {Horsley},\ and\ \citenamefont {Vos}}]{Barnes2020}%
  \BibitemOpen
  \bibfield  {author} {\bibinfo {author} {\bibfnamefont {W.~L.}\ \bibnamefont
  {Barnes}}, \bibinfo {author} {\bibfnamefont {S.~A.~R.}\ \bibnamefont
  {Horsley}}, \ and\ \bibinfo {author} {\bibfnamefont {W.~L.}\ \bibnamefont
  {Vos}},\ }\href@noop {} {\bibfield  {journal} {\bibinfo  {journal} {Journal
  of Optics}\ }\textbf {\bibinfo {volume} {7}} (\bibinfo {year}
  {2020})}\BibitemShut {NoStop}%
\bibitem [{\citenamefont {Markel}(1995)}]{Markel1995}%
  \BibitemOpen
  \bibfield  {author} {\bibinfo {author} {\bibfnamefont {V.~A.}\ \bibnamefont
  {Markel}},\ }\href {\doibase 10.1364/josab.12.001783} {\bibfield  {journal}
  {\bibinfo  {journal} {Journal of the Optical Society of America B}\ }\textbf
  {\bibinfo {volume} {12}},\ \bibinfo {pages} {1783} (\bibinfo {year}
  {1995})}\BibitemShut {NoStop}%
\bibitem [{\citenamefont {Merchiers}\ \emph {et~al.}(2007)\citenamefont
  {Merchiers}, \citenamefont {Moreno}, \citenamefont {Gonz{\'{a}}lez},\ and\
  \citenamefont {Saiz}}]{Merchiers2007}%
  \BibitemOpen
  \bibfield  {author} {\bibinfo {author} {\bibfnamefont {O.}~\bibnamefont
  {Merchiers}}, \bibinfo {author} {\bibfnamefont {F.}~\bibnamefont {Moreno}},
  \bibinfo {author} {\bibfnamefont {F.}~\bibnamefont {Gonz{\'{a}}lez}}, \ and\
  \bibinfo {author} {\bibfnamefont {J.~M.}\ \bibnamefont {Saiz}},\ }\href
  {\doibase 10.1103/PhysRevA.76.043834} {\bibfield  {journal} {\bibinfo
  {journal} {Physical Review A}\ }\textbf {\bibinfo {volume} {76}},\ \bibinfo
  {pages} {1} (\bibinfo {year} {2007})}\BibitemShut {NoStop}%
\bibitem [{\citenamefont {Bennett}\ and\ \citenamefont
  {Buhmann}(2020)}]{Bennett2019}%
  \BibitemOpen
  \bibfield  {author} {\bibinfo {author} {\bibfnamefont {R.}~\bibnamefont
  {Bennett}}\ and\ \bibinfo {author} {\bibfnamefont {S.~Y.}\ \bibnamefont
  {Buhmann}},\ }\href {http://iopscience.iop.org/10.1088/1367-2630/abac3a}
  {\bibfield  {journal} {\bibinfo  {journal} {New Journal of Physics}\ }
  (\bibinfo {year} {2020})}\BibitemShut {NoStop}%
\bibitem [{\citenamefont {Green}(2008)}]{Green2008}%
  \BibitemOpen
  \bibfield  {author} {\bibinfo {author} {\bibfnamefont {M.~A.}\ \bibnamefont
  {Green}},\ }\href@noop {} {\bibfield  {journal} {\bibinfo  {journal} {Solar
  Energy Materials and Solar Cells}\ }\textbf {\bibinfo {volume} {92}},\
  \bibinfo {pages} {1305} (\bibinfo {year} {2008})}\BibitemShut {NoStop}%
\bibitem [{\citenamefont {Bohren}\ and\ \citenamefont
  {Huffman}(2004)}]{Bohren2004}%
  \BibitemOpen
  \bibfield  {author} {\bibinfo {author} {\bibfnamefont {C.~F.}\ \bibnamefont
  {Bohren}}\ and\ \bibinfo {author} {\bibfnamefont {D.~R.}\ \bibnamefont
  {Huffman}},\ }\href@noop {} {\emph {\bibinfo {title} {Absorption and
  scattering of light by small particles}}}\ (\bibinfo  {publisher} {Wiley},\
  \bibinfo {address} {Weinheim},\ \bibinfo {year} {2004})\BibitemShut {NoStop}%
\bibitem [{\citenamefont {Evlyukhin}\ \emph {et~al.}(2010)\citenamefont
  {Evlyukhin}, \citenamefont {Reinhardt}, \citenamefont {Seidel}, \citenamefont
  {Luk'Yanchuk},\ and\ \citenamefont {Chichkov}}]{Evlyukhin2010}%
  \BibitemOpen
  \bibfield  {author} {\bibinfo {author} {\bibfnamefont {A.~B.}\ \bibnamefont
  {Evlyukhin}}, \bibinfo {author} {\bibfnamefont {C.}~\bibnamefont
  {Reinhardt}}, \bibinfo {author} {\bibfnamefont {A.}~\bibnamefont {Seidel}},
  \bibinfo {author} {\bibfnamefont {B.~S.}\ \bibnamefont {Luk'Yanchuk}}, \ and\
  \bibinfo {author} {\bibfnamefont {B.~N.}\ \bibnamefont {Chichkov}},\
  }\href@noop {} {\bibfield  {journal} {\bibinfo  {journal} {Physical Review
  B}\ }\textbf {\bibinfo {volume} {82}},\ \bibinfo {pages} {1} (\bibinfo {year}
  {2010})}\BibitemShut {NoStop}%
\bibitem [{\citenamefont {Pincus}(1970)}]{Pincus1970}%
  \BibitemOpen
  \bibfield  {author} {\bibinfo {author} {\bibfnamefont {M.}~\bibnamefont
  {Pincus}},\ }\href {\doibase 10.1287/opre.18.6.1225} {\bibfield  {journal}
  {\bibinfo  {journal} {Operations Research}\ }\textbf {\bibinfo {volume}
  {18}},\ \bibinfo {pages} {1225} (\bibinfo {year} {1970})}\BibitemShut
  {NoStop}%
\bibitem [{\citenamefont {Kraus}(1988)}]{antennas}%
  \BibitemOpen
  \bibfield  {author} {\bibinfo {author} {\bibfnamefont {J.~D.}\ \bibnamefont
  {Kraus}},\ }\href@noop {} {\emph {\bibinfo {title} {{Antennas}}}}\ (\bibinfo
  {publisher} {Tata McGraw-Hill},\ \bibinfo {address} {New York},\ \bibinfo
  {year} {1988})\BibitemShut {NoStop}%
\bibitem [{\citenamefont {Luneburg}\ and\ \citenamefont
  {Herzberger}(1964)}]{Luneburg1964}%
  \BibitemOpen
  \bibfield  {author} {\bibinfo {author} {\bibfnamefont {R.~K.}\ \bibnamefont
  {Luneburg}}\ and\ \bibinfo {author} {\bibfnamefont {M.}~\bibnamefont
  {Herzberger}},\ }\href@noop {} {\emph {\bibinfo {title} {Mathematical Theory
  of Optics}}}\ (\bibinfo  {publisher} {University of California Press},\
  \bibinfo {address} {Berkely and Los Angeles},\ \bibinfo {year}
  {1964})\BibitemShut {NoStop}%
\end{thebibliography}%

\end{document}